\documentclass[11pt]{article}

\usepackage{arxiv}
\usepackage[T1]{fontenc}
\usepackage{hyperref}
\usepackage{microtype}
\usepackage[backend=biber,sorting=none]{biblatex}
\usepackage{graphicx}

\usepackage{footnote}
\makesavenoteenv{tabular}
\makesavenoteenv{table}
\usepackage{enumitem}
\setlist[description]{leftmargin=24pt,labelindent=24pt}
\usepackage[title]{appendix}
\usepackage{titlesec}
\titlespacing\subsection{0pt}{12pt plus 4pt minus 2pt}{0pt plus 2pt minus 2pt}
\titlespacing\subsubsection{0pt}{12pt plus 4pt minus 2pt}{0pt plus 2pt minus 2pt}
\usepackage{fancyhdr}
\usepackage{color}
\newcommand{\chyperref}[2][ref]{{\color{blue} \hyperref[#1]{#2}}}

\title{Frontier AI Regulation:\goodbreak
	Managing Emerging Risks to Public Safety}
\renewcommand{\shorttitle}{Frontier AI Regulation}

\author{%
Markus~Anderljung$^{1,2*\dagger}$,
Joslyn~Barnhart$^{3**}$,
Anton~Korinek$^{4,5,1**\dagger}$,
Jade~Leung$^{6*}$,
Cullen~O'Keefe$^{6*}$,
Jess~Whittlestone$^{7**}$,
Shahar~Avin$^8$,
Miles Brundage$^6$,
Justin Bullock$^{9,10}$,
Duncan Cass-Beggs$^{11}$,
\newline
Ben Chang$^{12}$,
Tantum Collins$^{13,14}$,
Tim Fist$^2$,
Gillian Hadfield$^{15,16,17,6}$,
Alan Hayes$^{18}$,
Lewis Ho$^3$,
\newline
Sara Hooker$^{19}$,
Eric Horvitz$^{20}$,
Noam Kolt$^{15}$,
Jonas Schuett$^1$,
Yonadav Shavit$^{14***}$,
\newline
Divya Siddarth$^{21}$,
Robert Trager$^{1,22}$,
Kevin Wolf$^{18}$
}
\date{%
$^1$Centre for the Governance of AI,
$^2$Center for a New American Security,
$^3$Google DeepMind,
\newline
$^4$Brookings Institution,
$^5$University of Virginia, 
$^6$OpenAI,
$^7$Centre for Long-Term Resilience,
$^8$Centre for the Study of Existential Risk, University of Cambridge,
$^9$University of Washington,
$^{10}$Convergence Analysis,
$^{11}$Centre for International Governance Innovation,
$^{12}$The Andrew W. Marshall Foundation,
$^{13}$GETTING-Plurality Network, Edmond \&{} Lily Safra Center for Ethics,
$^{14}$Harvard University,
\newline
$^{15}$University of Toronto,
$^{16}$Schwartz Reisman Institute for Technology and Society,
$^{17}$Vector Institute,
$^{18}$Akin Gump Strauss Hauer \&{} Feld LLP,
$^{19}$Cohere For AI,
$^{20}$Microsoft,
$^{21}$Collective Intelligence Project,
$^{22}$University of California: Los Angeles
}

\newcommand\blfootnote[1]{%
  \begingroup
  \hypersetup{hidelinks}
  \renewcommand\thefootnote{}\footnote{#1}%
  \addtocounter{footnote}{-1}%
  \endgroup
}

\begin{document}
\maketitle
\blfootnote{Listed authors contributed substantive ideas and/or work to the white paper. Contributions include writing, editing, research, detailed feedback, and participation in a workshop on a draft of the paper. The first six authors are listed in alphabetical order, as are the subsequent 18. Given the size of the group, inclusion as an author does not entail endorsement of all claims in the paper, nor does inclusion entail an endorsement on the part of any individual's organization.}
\blfootnote{$^*$Significant contribution, including writing, research, convening, and setting the direction of the paper.}
\blfootnote{$^{**}$Significant contribution including editing, convening, detailed input, and setting the direction of the paper.}
\blfootnote{$^{***}$Work done while an independent contractor for OpenAI.}
\blfootnote{$^\dagger$Corresponding authors. Markus Anderljung (\texttt{markus.anderljung@governance.ai}) and Anton Korinek (\texttt{akorinek@brookings.edu}).}
\blfootnote{Cite as "Frontier AI Regulation: Managing Emerging Risks to Public Safety." Anderljung, Barnhart, Korinek, Leung, O'Keefe, \&{} Whittlestone, et al, 2023.}

\clearpage
\begin{abstract}
Advanced AI models hold the promise of tremendous benefits for humanity, but
society needs to proactively manage the accompanying risks. In this paper, we
focus on what we term “frontier AI” models — highly capable foundation models
that could possess dangerous capabilities sufficient to pose severe risks to
public safety.  Frontier AI models pose a distinct regulatory challenge:
dangerous capabilities can arise unexpectedly; it is difficult to robustly
prevent a deployed model from being misused; and, it is difficult to stop a
model’s capabilities from proliferating broadly. To address these challenges,
at least three building blocks for the regulation of frontier models are
needed: (1) standard-setting processes to identify appropriate requirements for
frontier AI developers, (2) registration and reporting requirements to provide
regulators with visibility into frontier AI development processes, and (3)
mechanisms to ensure compliance with safety standards for the development and
deployment of frontier AI models.  Industry self-regulation is an important
first step. However, wider societal discussions and government intervention
will be needed to create standards and to ensure compliance with them. We
consider several options to this end, including granting enforcement powers to
supervisory authorities and licensure regimes for frontier AI models. Finally,
we propose an initial set of safety standards. These include conducting
pre-deployment risk assessments; external scrutiny of model behavior; using
risk assessments to inform deployment decisions; and monitoring and responding
to new information about model capabilities and uses post-deployment. We hope
this discussion contributes to the broader conversation on how to balance
public safety risks and innovation benefits from advances at the frontier of AI
development.
\end{abstract}

\clearpage
\section*{Executive Summary}

The capabilities of today’s foundation models highlight both the promise and
risks of rapid advances in AI. These models have demonstrated significant
potential to benefit people in a wide range of fields, including education,
medicine, and scientific research. At the same time, the risks posed by
present-day models, coupled with forecasts of future AI progress, have
rightfully stimulated calls for increased oversight and governance of AI across
a range of policy issues. We focus on one such issue: the possibility that, as
capabilities continue to advance, new foundation models could pose severe risks
to public safety, be it via misuse or accident. Although there is ongoing
debate about the nature and scope of these risks, we expect that government
involvement will be required to ensure that such "frontier AI models” are
harnessed in the public interest.

Three factors suggest that frontier AI development may be in need of targeted
regulation: (1) Models may possess unexpected and difficult-to-detect dangerous
capabilities; (2) Models deployed for broad use can be difficult to reliably
control and to prevent from being used to cause harm; (3) Models may
proliferate rapidly, enabling circumvention of safeguards.

Self-regulation is unlikely to provide sufficient protection against the risks
from frontier AI models: government intervention will be needed. We explore
options for such intervention. These include:
\begin{quote}
\textbf{Mechanisms to create and update safety standards} for responsible
frontier AI development and deployment. These should be developed via
multi-stakeholder processes, and could include standards relevant to foundation
models overall, not exclusive to frontier AI. These processes should facilitate
rapid iteration to keep pace with the technology.

\textbf{Mechanisms to give regulators visibility} into frontier AI development,
such as disclosure regimes, monitoring processes, and whistleblower
protections. These equip regulators with the information needed to address the
appropriate regulatory targets and design effective tools for governing
frontier AI. The information provided would pertain to qualifying frontier AI
development processes, models, and applications.

\textbf{Mechanisms to ensure compliance with safety standards.} Self-regulatory
efforts, such as voluntary certification, may go some way toward ensuring
compliance with safety standards by frontier AI model developers. However, this
seems likely to be insufficient without government intervention, for example by
empowering a supervisory authority to identify and sanction non-compliance; or
by licensing the deployment and potentially the development of frontier AI.
Designing these regimes to be well-balanced is a difficult challenge; we should
be sensitive to the risks of overregulation and stymieing innovation on the one
hand, and moving too slowly relative to the pace of AI progress on the other.
\end{quote}

Next, we describe an initial set of safety standards that, if adopted,
would provide some guardrails on the development and deployment of frontier AI
models. Versions of these could also be adopted for current AI models to guard
against a range of risks. We suggest that at minimum, safety standards for
frontier AI development should include: 

\begin{quote}
\textbf{Conducting thorough risk assessments informed by evaluations of dangerous capabilities and controllability.} This would reduce the risk that
deployed models possess unknown dangerous capabilities, or behave unpredictably
and unreliably.

\textbf{Engaging external experts to apply independent scrutiny to
models.} External scrutiny of the safety and risk profile of models would both
improve assessment rigor and foster accountability to the public interest.

\textbf{Following standardized protocols for how frontier AI models
can be deployed based on their assessed risk.} The results from risk assessments
should determine whether and how the model is deployed, and what safeguards are
put in place. This could range from deploying the model without restriction to
not deploying it at all. In many cases, an intermediate option—deployment with
appropriate safeguards (e.g., more post-training that makes the model more
likely to avoid risky instructions)—may be appropriate.

\textbf{Monitoring and responding to new information on model
capabilities.} The assessed risk of deployed frontier AI models may change over
time due to new information, and new post-deployment enhancement techniques. If
significant information on model capabilities is discovered post-deployment,
risk assessments should be repeated, and deployment safeguards updated.
\end{quote}

Going forward, frontier AI models seem likely to warrant safety standards more
stringent than those imposed on most other AI models, given the prospective
risks they pose. Examples of such standards include: avoiding large jumps in
capabilities between model generations; adopting state-of-the-art alignment
techniques; and conducting pre-training risk assessments. Such practices are
nascent today, and need further development.

The regulation of frontier AI should only be one part of a broader policy
portfolio, addressing the wide range of risks and harms from AI, as well as
AI’s benefits. Risks posed by current AI systems should be urgently addressed;
frontier AI regulation would aim to complement and bolster these efforts,
targeting a particular subset of resource-intensive AI efforts. While we remain
uncertain about many aspects of the ideas in this paper, we hope it can
contribute to a more informed and concrete discussion of how to better govern
the risks of advanced AI systems while enabling the benefits of innovation to society.

\section*{Acknowledgements}

We would like to express our thanks to the people who have offered feedback and
input on the ideas in this paper, including Jon Bateman, Rishi
Bommasani, Will Carter, Peter Cihon, Jack Clark, John Cisternino, Rebecca
Crootof, Allan Dafoe, Ellie Evans, Marina Favaro, Noah Feldman, Ben Garfinkel,
Joshua Gotbaum, Julian Hazell, Lennart Heim, Holden Karnofsky, Jeremy Howard, Tim
Hwang, Tom Kalil, Gretchen Krueger, Lucy Lim, Chris Meserole,
Luke Muehlhauser, Jared Mueller, Richard Ngo, Sanjay Patnaik,
Hadrien Pouget, Gopal Sarma, Girish Sastry, Paul Scharre, Mike Selitto, Toby
Shevlane, Danielle Smalls, Helen Toner, and Irene Solaiman.

\clearpage
{\parskip=0em\tableofcontents}

\clearpage
\section{Introduction}

Responsible AI innovation can provide extraordinary benefits to society, such
as delivering medical \parencite{Moor2023, Lee2023, singhal2022large,
nori2023capabilities} and legal \parencite{Simshaw2022,Arbel2020,Kolt2022}
services to more people at lower cost, enabling scalable personalized education
\parencite{Khan2023}, and contributing solutions to pressing global challenges
like climate change \parencite{rolnick2019tackling, DeepMind2016,
erdinc2022derisking, Donti2021} and pandemic prevention
\parencite{Galetsi2022,danko2023challenges}. However, guardrails are necessary
to prevent the pursuit of innovation from imposing excessive negative
externalities on society. There is increasing recognition that government
oversight is needed to ensure AI development is carried out responsibly; we
hope to contribute to this conversation by exploring regulatory approaches to
this end.

In this paper, we focus specifically on the regulation of frontier AI models,
which we define as highly capable foundation models\footnote{Defined as: “any
model that is trained on broad data (generally using self-supervision at scale)
that can be adapted (e.g., fine-tuned) to a wide range of downstream tasks”
\parencite{bommasani2022opportunities}.} that could have dangerous capabilities
sufficient to pose severe risks to public safety and global security.  Examples
of such  dangerous capabilities include designing new biochemical weapons
\parencite{Urbina2022}, producing highly persuasive personalized
disinformation, and evading human control \parencite{ngo2023alignment,
Cohen2022, hendrycks2022unsolved, hendrycks2022xrisk,
carlsmith2022powerseeking, Russell:HumanCompatible, Christian:TheAlignmentProblem}.

In this paper, we first define frontier AI models and detail several policy
challenges posed by them. We explain why effective governance of frontier AI
models requires intervention throughout the models’ lifecycle, at the
development, deployment, and post-deployment stages.
Then, we describe approaches to regulating frontier AI models, including
building blocks of regulation such as the development of safety standards,
increased regulatory visibility, and ensuring compliance with safety standards.
We also propose a set of initial safety standards for frontier AI development
and deployment. We close by highlighting uncertainties and limitations for
further exploration.

\clearpage
\section{The Regulatory Challenge of Frontier AI Models}

\subsection{What do we mean by frontier AI models?}

For the purposes of this paper, we define “frontier AI models” as
highly capable foundation models\footnote{\cite{bommasani2022opportunities}
defines “foundation models” as “models (e.g., BERT, DALL-E, GPT-3) that are
trained on broad data at scale and are adaptable to a wide range of downstream
tasks.” See also \cite{EUDraft2023}.} that could exhibit sufficiently dangerous capabilities.
Such harms could take the form of significant physical harm or the disruption
of key societal functions on a global scale, resulting from intentional
misuse or accident \parencite{shevlane2023model, Zwetsloot2019}.
It would be prudent to assume that next-generation foundation models could
possess advanced enough capabilities to qualify as frontier AI models, given
both the \chyperref[sec:unexpected-capabilities-problem]{difficulty}
of predicting when sufficiently dangerous capabilities will
arise and the already significant capabilities of today’s models.

Though it is not clear where the line for “sufficiently dangerous capabilities”
should be drawn, examples could include:
\begin{itemize}
\item Allowing a non-expert to design and synthesize new biological or chemical weapons.\footnote{Such capabilities are starting to emerge. For example, a group of
researchers tasked a narrow drug-discovery system to identify maximally toxic
molecules. The system identified over 40,000 candidate molecules, including
both known chemical weapons and novel molecules that were predicted to be as or
more deadly \parencite{Urbina2022}.  Other researchers are warning that LLMs
can be used to aid in discovery and synthesis of compounds. One group attempted
to create an LLM-based agent, giving it access to the internet, code execution
abilities, hardware documentation, and remote control of an automated ‘cloud’
laboratory. They report finding that it in some cases the model was willing to outline
and execute on viable methods for synthesizing illegal drugs and chemical
weapons \parencite{boiko2023emergent}.}
\item Producing and propagating highly persuasive, individually tailored,
multi-modal disinformation with minimal user instruction.\footnote{%
Generative AI models may already be useful to generate material for
disinformation campaigns \parencite{Horvitz2022, goldstein2023generative,
Buchanan2021}.
It is possible that, in the future, models could possess additional
capabilities that could enhance the persuasiveness or dissemination of
disinformation, such as by making such disinformation more dynamic,
personalized, and multimodal; or by autonomously disseminating such
disinformation through channels that enhance its persuasive value, such as
traditional media.}
\item Harnessing unprecedented offensive cyber capabilities that could cause
catastrophic harm.\footnote{AI systems are already helpful in writing and
debugging code, capabilities that can also be applied to software vulnerability
discovery. There is potential for significant harm via automation of
vulnerability discovery and exploitation. However, vulnerability discovery
could ultimately benefit cyberdefense more than -offense, provided defenders
are able to use such tools to identify and patch vulnerabilities more
effectively than attackers can find and exploit them
\parencite{poldrack2023aiassisted, Lohn2022}.}
\item Evading human control through means of deception and obfuscation.\footnote{If future AI systems develop the ability and the propensity to
deceive their users, controlling their behavior could be extremely challenging.
Though it is unclear whether models will trend in that direction, it seems rash
to dismiss the possibility and some argue that it might be the default outcome
of current training paradigms \parencite{ngo2023alignment, Cohen2022,
hendrycks2022xrisk, carlsmith2022powerseeking, Russell:HumanCompatible,
Christian:TheAlignmentProblem}.}
\end{itemize}

This list represents just a few salient possibilities; the possible future
capabilities of frontier AI models remains an important area of inquiry.

\begin{figure}
\begin{center}
\includegraphics[width=\textwidth]{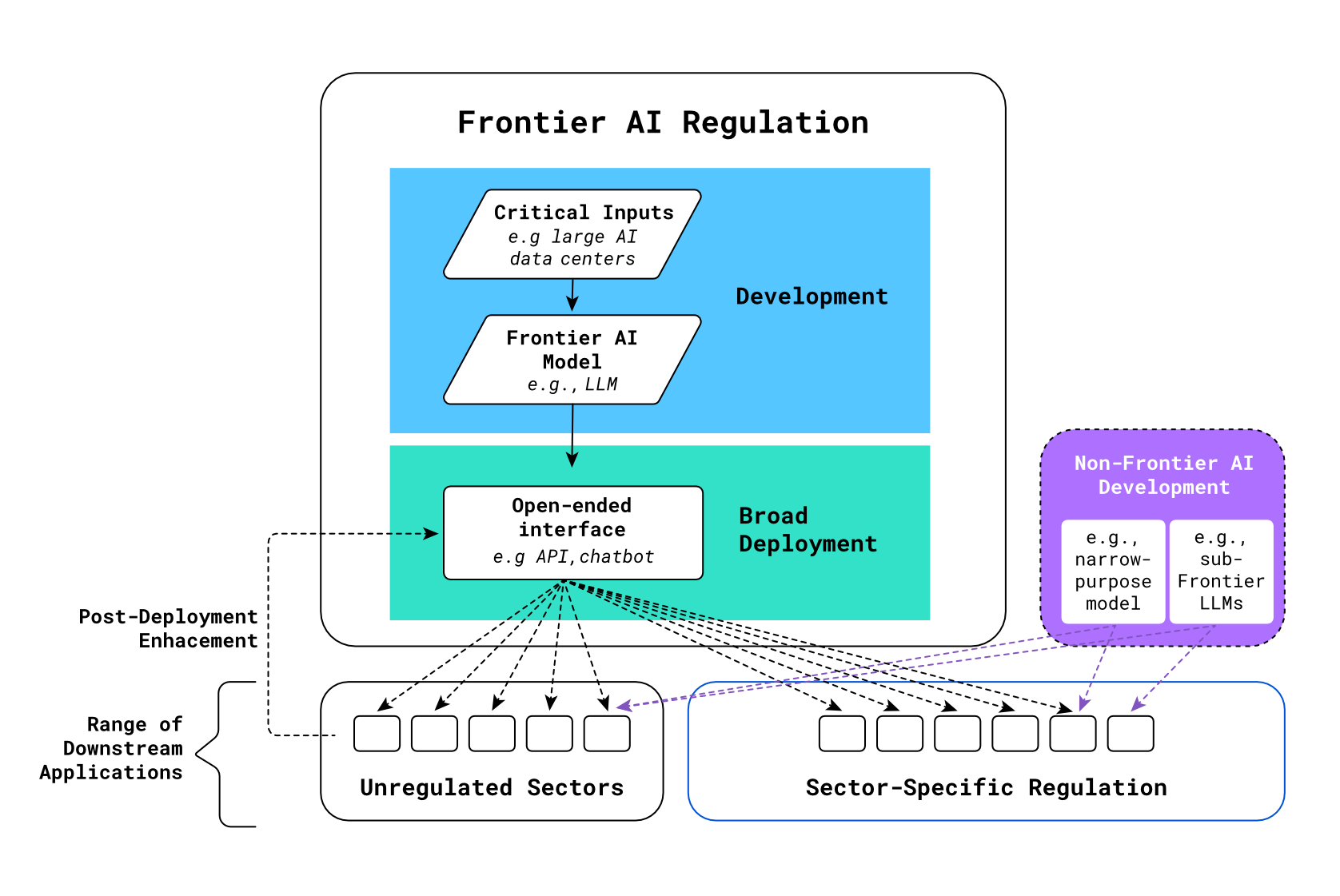}
\end{center}
\caption{Example frontier AI lifecycle.}
\label{figure:lifecycle}
\end{figure}

Foundation models, such as large language models (LLMs), are trained on large,
broad corpora of natural language and other text (e.g., computer code), usually
starting with the simple objective of predicting the next “token”.\footnote{A
token can be thought of as a word or part of a word
\parencite{Microsoft:token}.} This relatively simple approach produces models
with surprisingly broad capabilities.\footnote{For example, LLMs achieve
state-of-the-art performance in diverse tasks such as question answering,
translation, multi-step reasoning, summarization, and code completion, among
others \parencite{Radford2019, brown2020language, openai2023gpt4,
chowdhery2022palm}. Indeed, the term “LLM” is already becoming outdated, as
several leading “LLMs” are in fact multimodal (e.g., possess visual
capabilities) \parencite{openai2023gpt4, alayrac2022flamingo}.}  These models
thus possess more general-purpose functionality\footnote{We intentionally avoid
using the term “general-purpose AI” to avoid confusion with the use of that
term in the EU AI Act and other legislation. Frontier AI systems are a related
but narrower class of AI systems with general-purpose functionality, but whose
capabilities are relatively advanced and novel.} than many other classes of AI
models, such as the recommender systems used to suggest Internet videos or
generative AI models in narrower domains like music.  Developers often make
their models available through “broad deployment” via sector-agnostic platforms
such as APIs, chatbots, or via open-sourcing.\footnote{We use “open-source” to
mean “open release:” that is a model being made freely available online, be it
with a license restricting what the system can be used for. An example of such
a license is the Responsible AI License. Our usage of “open-source” differs
from how the term is often used in computer science which excludes instances of
license requirements, though is closer to how many other communities understand
the term \parencite{licensesai,opensource}.} This means that they can be
integrated in a large number of diverse downstream applications, possibly
including safety-critical sectors (illustrated in
\chyperref[figure:lifecycle]{Figure~\ref{figure:lifecycle}}).

A number of features of our definition are worth highlighting. In focusing on
\emph{foundation models} which could have dangerous, emergent capabilities, our
definition of frontier AI excludes narrow models, even when these models could
have sufficiently dangerous capabilities.\footnote{However, if a foundation
model could be fine-tuned and adapted to pose severe risk to public safety via
capabilities in some narrow domain, it would count as a “frontier AI.”} For
example, models optimizing for the toxicity of compounds \parencite{Urbina2022}
or the virulence of pathogens could lead to intended (or at least foreseen)
harms and thus may be more appropriately covered with more targeted
regulation.\footnote{Indeed, intentionally creating dangerous narrow models
should already be covered by various laws and regulators. To the extent that it
is not clearly covered, modification of those existing laws and regulations
would be appropriate and urgent. Further, the difference in mental state of the
developer makes it much easier to identify and impose liability on developers
of narrower dangerous models.}

Our definition focuses on models that \emph{could} — rather than just those
that \emph{do} — possess dangerous capabilities, as many of the practices we
propose apply before it is known that a model has dangerous capabilities. One
approach to identifying models that could possess such capabilities is focusing on
foundation models that advance the state-of-the-art of foundation model
capabilities. While currently deployed foundation models pose risks
\parencite{bommasani2022opportunities, Bender2021}, they do not yet appear to
possess dangerous capabilities that pose severe risks to public safety as we
have defined them.\footnote{In some cases, these have been explicitly tested
for \parencite{openai:system-card}.} Given both our \chyperref[sec:unexpected-capabilities-problem]{inability to reliably predict}
what models will have sufficiently dangerous capabilities and the
already significant capabilities today’s models possess, it would be prudent
for regulators to assume that next-generation state-of-the-art foundation
models \emph{could} possess advanced enough capabilities to warrant
regulation.\footnote{We think it is prudent to anticipate that foundation
models’ capabilities may advance much more quickly than many expect, as has
arguably been the case for many AI capabilities: “[P]rogress on ML benchmarks
happened significantly faster than forecasters expected. But forecasters
predicted faster progress than I did personally, and my sense is that I expect
somewhat faster progress than the median ML researcher does.”
\parencite{Steinhardt2023}; See \cite{zhang2022forecasting} at 9;
\cite{hoffmann2022training} at 11 (Chinchilla and Gopher surpassing forecaster
predictions for progress on MMLU); \cite{openai2023gpt4} (GPT-4 surpassing
Gopher and Chinchilla on MMLU, also well ahead of forecaster predictions);
\cite{Caplan:new, Caplan:retakes, Metaculus:Go, Metaculus:programs}.} An
initial way to identify potential state-of-the-art foundation models could be
focusing on models trained using above some very large amount of computational
resources.\footnote{Perhaps more than any model that has been trained to date.
Estimates suggest that 1E26 floating point operations (FLOP) would meet this
criteria \parencite{OWID:computation}.}

Over time, the scope of frontier AI should be further refined. The scope
should be sensitive to features other than compute; state-of-the-art
performance can be achieved by using high quality data and new algorithmic
insights. Further, as systems with sufficiently dangerous capabilities are
identified, it will be possible to identify training runs that are likely to
produce such capabilities despite not achieving state-of-the-art performance.

We acknowledge that our proposed definition is lacking in sufficient precision
to be used for regulatory purposes and that more work is required to fully
assess the advantages and limitations of different approaches. Further, it is
not our role to determine exactly what should fall within the scope of the
regulatory proposals outlined – this will require more analysis and input from
a wider range of actors. Rather, the aim of this paper is to present a set of
initial proposals which we believe should apply to at least some subset of AI
development. We provide a more detailed description of alternative approaches
and the general complexity of defining “frontier AI” in \chyperref[appendix:a]{Appendix~A}.

\subsection{The Regulatory Challenge Posed by Frontier AI}

There are many regulatory questions related to the widespread use of AI
\parencite{bommasani2022opportunities}. This paper focuses on a specific subset
of concerns: the possibility that continued development of increasingly capable
foundation models could lead to dangerous capabilities sufficient to pose risks
to public safety at even greater severity and scale than is possible with
current computational systems \parencite{shevlane2023model}.

Many existing and proposed AI regulations focus on the context in which AI
models are deployed, such as high-risk settings like law enforcement and
safety-critical infrastructure.  These proposals tend to favor sector-specific
regulations models.\footnote{This could look like imposing new requirements for
AI models used in high-risk industries and modifying existing regulations to
account for new risks from AI models. See \parencite{EUDraft2023,DCIAct,AAact,
ActionPlan2021,Consumer2022,LinaKhan2023}.} For frontier AI development,
sector-specific regulations can be valuable, but will likely leave a subset of
the high severity and scale risks unaddressed.

Three core problems shape the regulatory challenge posed by frontier AI models:
\begin{quote}
\textbf{\chyperref[sec:unexpected-capabilities-problem]{The Unexpected Capabilities Problem.}}
Dangerous capabilities can arise
unpredictably and undetected, both during development and after deployment.

\textbf{\chyperref[sec:deployment-safety-problem]{The Deployment Safety Problem.}}
Preventing deployed AI models from causing
harm is a continually evolving challenge.

\textbf{\chyperref[sec:proliferation-problem]{The Proliferation Problem.}}
Frontier AI models can proliferate rapidly,
making accountability difficult.
\end{quote}

These problems make the regulation of frontier AI models fundamentally
different from the regulation of other software, and the majority of other AI
models. The \emph{Unexpected Capabilities Problem} implies that frontier AI
models could have unpredictable or undetected dangerous capabilities that
become accessible to downstream users who are difficult to predict beforehand.
Regulating easily identifiable users in a relatively small set of
safety-critical sectors may therefore fail to prevent those dangerous
capabilities from causing significant harm.\footnote{This is especially true for
downstream bad actors (e.g., criminals, terrorists, adversary nations), who
will tend not to be as regulable as the companies operating in domestic
safety-critical sectors.}

\emph{The Deployment Safety Problem} adds an additional layer of difficulty.
Though many developers implement measures intended to prevent models from
causing harm when used by downstream users, these may not always be foolproof,
and malicious users may constantly be attempting to evolve their attacks.
Furthermore, the Unexpected Capabilities Problem implies that the developer may
not know of all of the harms from frontier models that need to be guarded
against during deployment. This amplifies the difficulty of the Deployment
Safety Problem: deployment safeguards should address not only known dangerous
capabilities, but have the potential to address unknown ones too.

The \emph{Proliferation Problem} exacerbates the regulatory challenge. Frontier
AI models may be open-sourced, or become a target for theft by adversaries. To
date, deployed models also tend to be reproduced or iterated on within several
years. If, due to the Unexpected Capabilities Problem, a developer (knowingly
or not) develops and deploys a model with dangerous capabilities, the
Proliferation Problem implies that those capabilities could quickly become
accessible to unregulable actors like criminals and adversary governments.

Together, these challenges show that adequate regulation of frontier AI should
intervene throughout the frontier AI lifecycle, including during development,
general-purpose deployment, and post-deployment enhancements.

\subsubsection{The Unexpected Capabilities Problem: Dangerous Capabilities Can
Arise Unpredictably and Undetected}\label{sec:unexpected-capabilities-problem}

Improvements in AI capabilities can be unpredictable, and are often difficult
to fully understand without intensive testing. Regulation that does not require
models to go through sufficient testing before deployment may therefore fail to
reliably prevent deployed models from posing severe risks.\footnote{This
challenge also exacerbates the Proliferation Problem: we may not know how
important nonproliferation of a model is until after it has already been
open-sourced, reproduced, or stolen.}

Overall AI model performance\footnote{Measured by loss: essentially the error
rate of an AI model performs on its training objective. We acknowledge that
this is not a complete measure of model performance by any means.} has tended to
improve smoothly with additional compute, parameters, and data.\footnote{See
\parencite{kaplan2020scaling,henighan2020scaling,hoffmann2022training,
Villalobos2023,hestness2017deep}
However, there are tasks for which scaling leads to worse performance
\parencite{mckenzie2023inverse, perez2022discovering, koralus2023humans},
though further scaling has overturned some of these findings,
\parencite{openai2023gpt4}. See also \chyperref[appendix:b]{Appendix~B}.} However, specific
capabilities can significantly improve quite suddenly in general-purpose models
like LLMs (see \chyperref[figure:emerge]{Figure~2}).
Though debated (see \chyperref[appendix:b]{Appendix~B}),
this phenomenom has been repeatedly
observed in multiple LLMs with capabilities as diverse as modular arithmetic,
unscrambling words, and answering questions in Farsi
\parencite{wei2022emergent, Wei2022, bowman2023things, Wei2023}.\footnote{For a
treatment of recent critiques of the claim that AI models exhibit emergent
capabilities, see Appendix~B.} Furthermore, given the vast set of possible
tasks a foundation model could excel at, it is nearly impossible to
exhaustively test for them \parencite{bommasani2022opportunities,
shevlane2023model}

\begin{figure}
\begin{center}
\includegraphics[width=.9\textwidth]{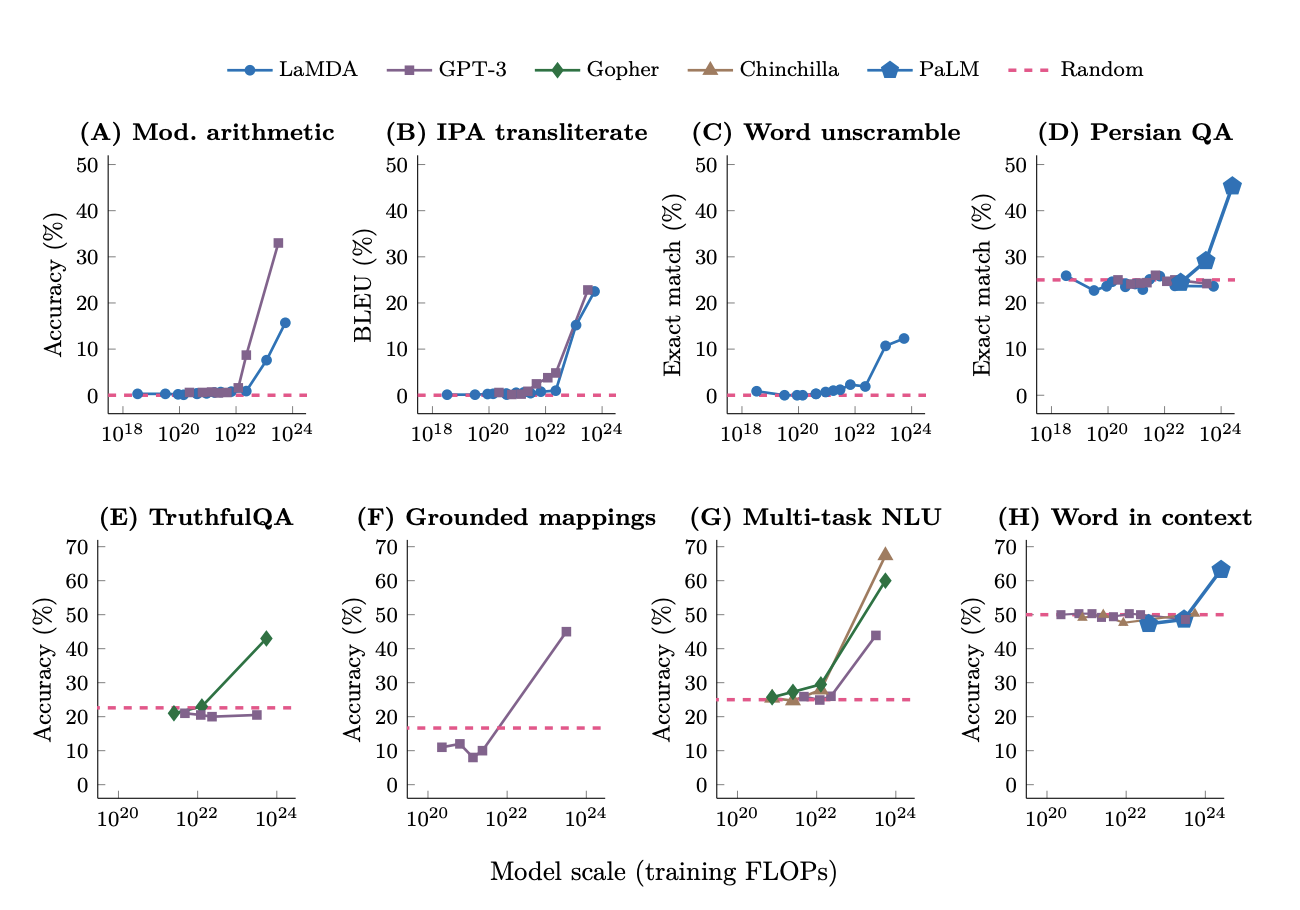}
\end{center}
\caption[Certain capabilities seem to emerge suddenly.]{Certain capabilities
seem to emerge suddenly\footnotemark}
\label{figure:emerge}
\end{figure}

\footnotetext{Chart from \cite{wei2022emergent}. But see
\cite{schaeffer2023emergent} for a skeptical view on emergence. For a response
to the skeptical view, see~\cite{Wei2023} and Appendix~B.}

Post-deployment enhancements — modifications made to AI models after their
initial deployment — can also cause unaccounted-for capability jumps. For
example, a key feature of many foundation models like LLMs is that they can be
fine-tuned on new data sources to enhance their capabilities in targeted
domains.  AI companies often allow customers to fine-tune foundation models on
task-specific data to improve the model’s performance on that task
\parencite{anthropic, OpenAI:fine-tuning, ai21, cohere}.
This could effectively expand the scope of capability concerns of a particular
frontier AI model.  Models could also be improved via “online” learning, where
they continuously learn from new data \parencite{hoi2018online,Parisi2019}.

To date, iteratively deploying models to subsets of users has been a key
catalyst for understanding the outer limits of model capabilities and
weaknesses.\footnote{Dario Amodei, CEO of Anthropic: ``You have to deploy it to
a million people before you discover some of the things that it can do\dots''
\parencite{WPost2023}. “We work hard to prevent foreseeable risks before
deployment, however, there is a limit to what we can learn in a lab.
Despite extensive research and testing, we cannot predict all of the beneficial
ways people will use our technology, nor all the ways people will abuse it.
That’s why we believe that learning from real-world use is a critical component
of creating and releasing increasingly safe AI systems over time”
\parencite{OpenAI:safety}.} For example, model users have demonstrated
significant creativity in eliciting new capabilities from AI models, exceeding
developers’ expectations of model capabilities. Users continue to discover
prompting techniques that significantly enhance the model’s performance, such
as by simply asking an LLM to reason step-by-step
\parencite{wei2023chainofthought}. This has been described as the “capabilities
overhang” of foundation models \parencite{Clark2022}. Users also discover new
failure modes for AI systems long after their initial deployment. For example,
one user found that the string “ solidgoldmagikarp” caused GPT-3 to malfunction
in a previously undocumented way, years after that model was first deployed
\parencite{SolidGoldMagikarp}.

Much as a carpenter’s overall capabilities will vary with the tools she has
available, so too might an AI model’s overall capabilities vary depending on
the tools it can use. LLMs can be taught to use, and potentially create,
external tools like calculators and search engines
\parencite{OpenAI:plugins,schick2023toolformer,cai2023large}.
Some models are also being trained to directly use general-purpose mouse and
keyboard interfaces \parencite{Adept,AutoGPT}. See more examples in \chyperref[table:post-deployment]{Table~1}.
As the available tools improve,
so can the overall capabilities of the total model-tool system, even if the
underlying model is largely unchanged.\footnote{Right now, most tools that AI
models can use were originally optimized for use by people. As model-tool
interactions become more economically important, however, companies may develop
tools optimized for use by frontier AI models, accelerating capability
improvements.}

\begin{table}
\begin{center}
\begin{tabular}{l|p{4.5cm}|p{5.5cm}}
	\textbf{Technique} & \textbf{Description} & \textbf{Example}\\
	\hline
	Fine-tuning & Improving foundation model performance by updating model
	weights with task-specific data. & Detecting propaganda by fine-tuning
	a pre-trained LLM on a labeled dataset of common propaganda tactics
	\parencite{Yoosuf2019}.\\
	Chain-of-thought prompting \parencite{wei2023chainofthought} &
	Improving LLM problem-solving capabilities by telling the model to
	think through problems step by step. & Adding a phrase such as “Let’s
	think step by step” after posing a question to the model
	\parencite{kojima2023large}.\\
	External tool-use & Allow the model to use external tools when figuring
	out how to answer user queries. & A model with access to a few simple
	tools (e.g., calculator, search engine) and a small number of examples
	performs much better than an unaided model.\footnote{See
	\parencite{schick2023toolformer}. Early research also suggests LLMs can
	be used to create tools for their own use \parencite{cai2023large}.}\\
	Automated prompt engineering \parencite{zhou2023large} & Using LLMs to
	generate and search over novel prompts that can be used to elicit
	better performance on a task. & To generate prompts for a task, an LLM
	is asked something akin to: “I gave a friend instructions and he
	responded in this way for the given inputs: [Examples of inputs and
	outputs of the task] The instruction was:”\\
	Foundation model programs \parencite{schlag2023large} & Creation of
	standardized means of integrating foundation models into more complex
	programs. & Langchain: “a framework for developing applications powered
	by language models.” \parencite{LangChain,AutoGPT}
\end{tabular}
\end{center}
\caption{Some known post-deployment techniques for unlocking new AI
capabilities.}
\label{table:post-deployment}
\end{table}

In the long run, there are even more worrisome possibilities. Models behaving
differently in testing compared to deployment is a known phenomenon in the
field of machine learning, and is particularly worrisome if unexpected and
dangerous behaviors first emerge “in the wild” only once a frontier model is
deployed \parencite{turner2023optimal, krakovna2023powerseeking,
hubinger2021risks}.

\subsubsection{The Deployment Safety Problem: Preventing Deployed AI Models
from Causing Harm is Difficult}\label{sec:deployment-safety-problem}

In general, it is difficult to precisely specify what we want deep
learning-based AI models to do and to ensure that they behave in line with
those specifications. Reliably controlling powerful AI models’ behavior, in
other words, remains a largely unsolved technical problem
\parencite{hendrycks2022unsolved, ngo2023alignment, amodei2016concrete,
wolf2023fundamental, bowman2023things} and the subject of ongoing research.

Techniques to “bake in” misuse prevention features at the model level, such
that the model reliably rejects or does not follow harmful instructions, can
effectively mitigate these issues, but adversarial users have still found ways
to circumvent these safeguards in some cases. One technique for circumvention
has been prompt injection attacks, where attackers disguise input text as
instructions from the user or developer to overrule restrictions provided to or
trained into the model. For example, emails sent to an LLM-based email
assistant could contain text constructed to look to the user as benign, but to
the LLM contains instructions to exfiltrate the user’s data (which the LLM
could then follow).\footnote{For additional examples, see \cite{Willison2023}.}
Other examples include “jailbreaking” models by identifying prompts that cause
a model to act in ways discouraged by their developers
\parencite{Venuto2023, PromptReport, Metz2023}.  Although progress is
being made on such issues \parencite{bai2022constitutional, pan2023rewards,
Venuto2023, openai:system-card}, it is unclear that we will be able to reliably
prevent dangerous capabilities from being used in unintended or undesirable
ways in novel situations; this remains an open and fundamental technical
challenge.

A major consideration is that model capabilities can be employed for  both
harmful and beneficial uses:\footnote{Nearly all attempts to stop bad or
unacceptable uses of AI also hinder positive uses, creating a \emph{Misuse-Use
Tradeoff} \parencite{anderljung2023protecting}.} the harmfulness of an AI
model’s action may depend almost entirely on context that is not visible
during model development.
For example, copywriting is helpful when a company uses it to generate internal
communications, but harmful when propagandists use it to generate or amplify
disinformation.  Use of a text-to-image model to modify a picture of someone
may be used with their consent as part of an art piece, or without their
consent as a means of producing disinformation or harassment.

\subsubsection{The Proliferation Problem: Frontier AI Models Can Proliferate
Rapidly}\label{sec:proliferation-problem}

The most advanced AI models cost tens of millions of dollars to
create.\footnote{Though there are no estimates on the total cost of producing a
frontier model, there are estimates of the cost of the compute used to train
models \parencite{Heim2022,Sevilla2022,Cottier2022}} However, using the
trained model (i.e., “inference”) is vastly cheaper.\footnote{Some impressive
models can run on a offline portable device; see
\cite{Orhon2022,Willison:LLaMa,gpt4all,chen2023speed}.} Thus, a much wider
array of actors will have the resources to misuse frontier AI models than have
the resources to create them. Those with access to a model with dangerous
capabilities could cause harm at a significant scale, by either misusing the
model themselves, or passing it on to actors who will misuse it.\footnote{Though
advanced computing hardware accessed via the cloud tends to be needed to use
frontier models. They can seldom be run on consumer-grade hardware.} We describe some examples of proliferation in \chyperref[table:risk]{Table~2}.

Currently, state-of-the-art AI capabilities can proliferate soon after
development. One mechanism for proliferation is open-sourcing. At present,
proliferation via open-sourcing of advanced AI models is common\footnote{ For an overview of considerations in how to release powerful AI models, see
\cite{solaiman2019release, solaiman2023gradient, shevlane2022structured,
Nature2021, ovadya2019reducing, Sastry2021}.} \parencite{Leahy2021, BLOOM,
HuggingFace}  and usually unregulated. When models are open-sourced, obtaining
access to their capabilities becomes much easier: all internet users could copy
and use them, provided access to appropriate computing resources. Open-source
AI models can provide major economic utility by driving down the cost of
accessing state-of-the-art AI capabilities. They also enable academic research
on larger AI models than would otherwise be practical, which improves the
public’s ability to hold AI developers accountable. We believe that
open-sourcing AI models can be an important public good. However, frontier AI
models may need to be handled more restrictively than their smaller, narrower,
or less capable counterparts. Just as cybersecurity researchers embargo
security vulnerabilities to give the affected companies time to release a
patch, it may be prudent to avoid potentially dangerous capabilities of
frontier AI models being open sourced until safe deployment is demonstrably
feasible. 

Other vectors for proliferation also imply increasing risk as capabilities
advance. For example, though models that are made available via APIs
proliferate more slowly, newly announced results are commonly reproduced or
improved upon\footnote{Below, we use “reproduction” to mean some other actor
producing a model that reaches at least the same performance as an existing
model.} within 1-2 years of the initial release. Many of the most capable
models use simple algorithmic techniques and freely available data, meaning
that the technical barriers to reproduction can often be low.\footnote{Projects
such as OpenAssistant \parencite{OpenAssistant} attempt to reproduce the
functionality of ChatGPT; and alpaca \parencite{alpaca} uses OpenAI’s
text-davinci-003 model to train a new model with similar capabilities. For an
overview, see \cite{zhao2023survey}.}

Proliferation can also occur via theft. The history of cybersecurity is replete
with examples of actors ranging from states to lone cybercriminals compromising
comparably valuable digital assets \parencite{Maness2022, Carnegie, CSIS,
Schmidt2015, Buchanan2017}. Many AI developers take significant measures to
safeguard their models. However, as AI models become more useful in
strategically important contexts and the difficulties of producing the most
advanced models increase, well-resourced adversaries may launch increasingly
sophisticated attempts to steal them \parencite{Hannas2019, NCSC2021}.
Importantly, theft is feasible before deployment.

\begin{figure}
\centering\includegraphics[width=\textwidth]{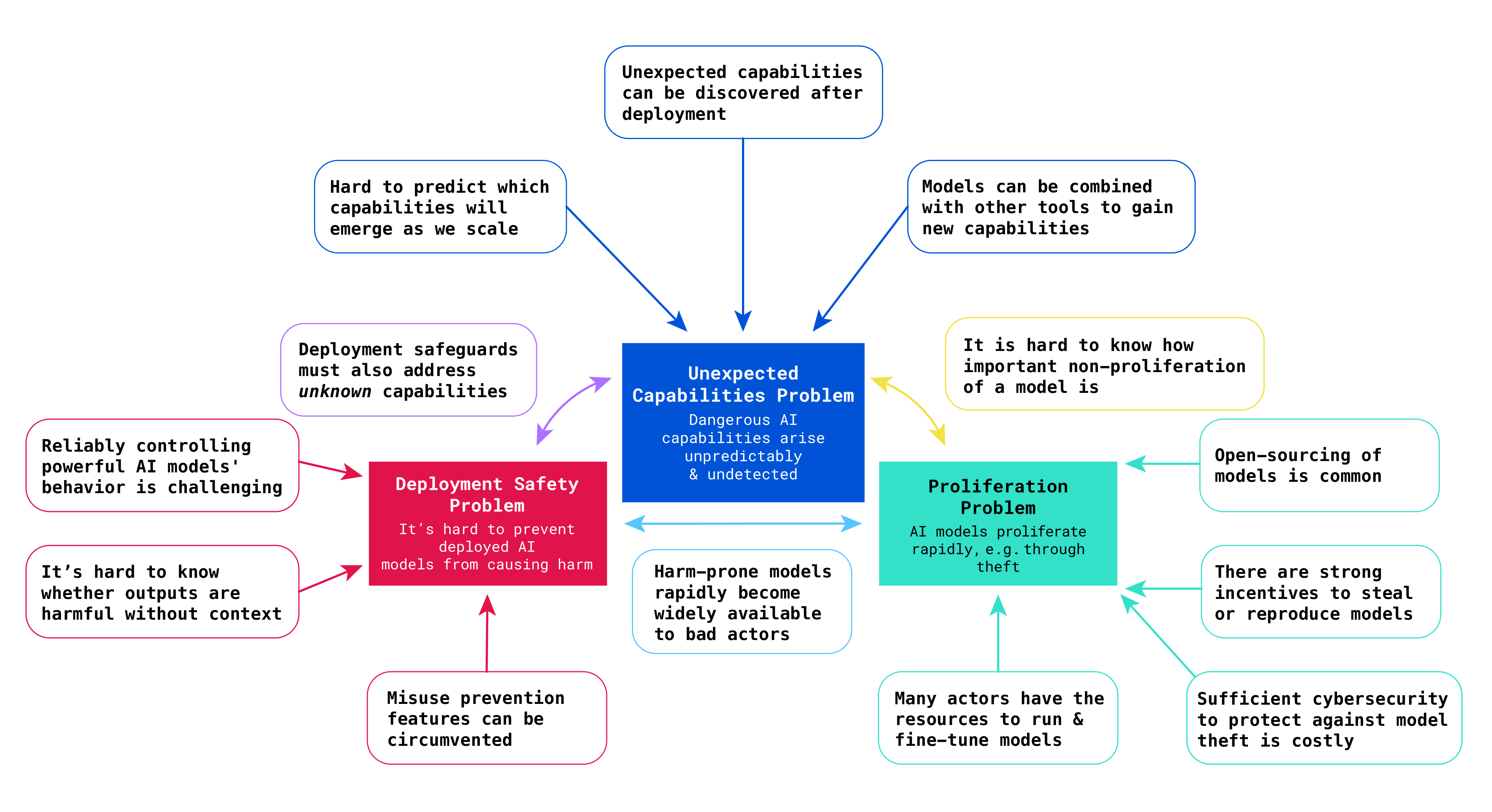}
\caption{Summary of the three regulatory challenges posed by frontier AI.}
\label{figure:challenges}
\end{figure}

The interaction and causes of the three regulatory challenges posed by frontier
AI are summarized in \chyperref[figure:challenges]{Figure~\ref{figure:challenges}}.

\begin{table}
\centering
\begin{tabular}{l|l|l}
\hline\textbf{Original Model} & \textbf{Subsequent Model} & \textbf{Time to
Proliferate}\footnote{The examples listed here are not necessarily the earliest
instances of proliferation.}\\
\hline\multicolumn2{l|}{StyleGAN} & Immediate\\
\hline\multicolumn3{p{\textwidth}}{StyleGAN is a model by NVIDIA that
generates photorealistic human faces using generative adversarial networks
(GANs) \parencite{stylegan}. NVIDIA first published about StyleGAN in December
2018 \parencite{karras2019stylebased} and open-sourced the model in February
2019. Following open-sourcing StyleGAN, sample images went viral through sites
such as \texttt{thispersondoesnotexist.com}
\parencite{Metz:fake-images,thispersondoesnotexist}. Fake social media accounts
using pictures from StyleGAN were discovered later that year
\parencite{Gallagher:facebook,Bond:fake-faces}.}\\
\hline AlphaFold 2 & OpenFold & $\sim$2 years\\
\hline\multicolumn3{p{\textwidth}}{In November 2020, DeepMind announced
AlphaFold~2 \parencite{AlphaFold}. It was “the first computational method that
can regularly predict protein structures with atomic accuracy even in cases in
which no similar structure is known” \parencite{Jumper2021}: a major advance in
the biological sciences. In November~2022, a diverse group of researchers
reproduced and open-sourced a similarly capable model named OpenFold
\parencite{Ahdritz2022}. OpenFold used much less data to train than AlphaFold
2, and could be run much more quickly and easily \parencite{Ahdritz2022}.}\\
\hline GPT-3 & Gopher & $\sim$7 months\\
\hline\multicolumn3{p{\textwidth}}{OpenAI announced GPT-3, an LLM, in May
2020 \parencite{brown2020language}. In December~2021, DeepMind announced
Gopher, which performed better than GPT-3 across a wide range of benchmarks.
However, the Gopher model card suggests that the model was developed significantly earlier,
seven months after the GPT-3 announcement, in December~2020
\parencite{rae2022scaling}.}\\
\hline\multicolumn2{l|}{LLaMa} & $\sim$1 week\\
\hline\multicolumn3{p{\textwidth}}{In February 2023, Meta AI announced LLaMa,
an LLM \parencite{LLaMA}. LLaMa was not open-sourced, but researchers could
apply for direct access to model weights \parencite{LLaMA}. Within a week,
various users had posted these weights on multiple websites, violating the
terms under which the weights were distributed \parencite{TheBatch2023}.}\\
\hline ChatGPT & Alpaca & $\sim$3 months\\
\hline\multicolumn3{p{\textwidth}}{In March 2023, researchers from Stanford
University used sample completions from OpenAI’s text-davinci-003 to fine-tune
LLaMa in an attempt to recreate ChatGPT using less than \$600.\footnote{Note
that the original paper and subsequent research suggests this method fails to
match the capabilities of the larger model
\parencite{alpaca,gudibande2023false}.} Their model was subsequently taken
offline due to concerns about cost and safety \parencite{Quach2023}, though the
code and documentation for replicating the model is available on GitHub
\parencite{tatsu-lab:alpaca}.}\\
\hline
\end{tabular}
\caption{Examples of AI proliferation: these are not necessarily typical, and
some of these examples may be beneficial or benign, yet they demonstrate the
consistent history of AI capabilities proliferating after their initial
deployment}
\end{table}

% \clearpage
\section{Building Blocks for Frontier AI Regulation}

The three problems described above imply that serious risks may emerge during
the development and deployment of a frontier AI model, not just when it is used
in safety-critical sectors. Regulation of frontier AI models, then, must
address the particular shape of the regulatory challenge: the potential
unexpected dangerous capabilities; difficulty of deploying AI models safely;
and the ease of proliferation.

In this section, we outline potential building blocks for the regulation of
frontier AI. In the \chyperref[sec:initial-standards]{next section}, we describe a set of initial safety standards
for frontier AI models that this regulatory regime could ensure developers
comply with. 

Much of what we describe could be helpful frameworks for understanding how to
address the range of challenges posed by current AI models. We also acknowledge
that much of the discussion below is most straightforwardly applicable to the
context of the United States. Nevertheless, we hope that other jurisdictions
could benefit from these ideas, with appropriate modifications. 

A regulatory regime for frontier AI would likely need to include a number of
building blocks: 

\begin{quote}
\textbf{\chyperref[sec:institutionalize]{Mechanisms for development of frontier AI safety standards}}
particularly
via expert-driven multi-stakeholder processes, and potentially coordinated by
governmental bodies. Over time, these standards could become enforceable legal
requirements to ensure that frontier AI models are being developed safely.

\textbf{\chyperref[sec:increase-visibility]{Mechanisms to give regulators visibility}}
into frontier AI development,
such as disclosure regimes, monitoring processes, and whistleblower protection.
These equip regulators with the information needed to address the appropriate
regulatory targets and design effective tools for governing frontier AI.

\textbf{\chyperref[sec:ensure-compliance]{Mechanisms to ensure compliance with safety standards}}
including
voluntary self-certification schemes, enforcement by supervisory authorities,
and licensing regimes. While self-regulatory efforts, such as voluntary
certification, may go some way toward ensuring compliance, this seems likely to
be insufficient for frontier AI models.
\end{quote}

Governments could encourage the development of standards and consider
increasing regulatory visibility today; doing so could also address potential
harms from existing systems. We expand on the conditions under which more
stringent tools like enforcement by supervisory authorities or licensing may be
warranted \chyperref[sec:license]{below}.

Regulation of frontier AI should also be complemented with efforts to reduce
the harm that can be caused by various dangerous capabilities. For example, in
addition to reducing frontier AI model usefulness in designing and producing
dangerous pathogens, DNA synthesis companies should screen for such worrying
genetic sequences \parencite{soice2023large, anderljung2023protecting}. While
we do not discuss such efforts to harden society against the proliferation of
dangerous capabilities in this paper, we welcome such efforts from others.

\subsection{Institutionalize Frontier AI Safety Standards Development}
\label{sec:institutionalize}

Policymakers should support and initiate sustained, multi-stakeholder processes
to develop and continually refine the safety standards that developers of
frontier AI models may be required to adhere to. To seed these processes, AI
developers, in partnership with civil society and academia, can pilot practices
that improve safety during development and deployment
\parencite{Google:ResponsibleAI, Cohere2022, Microsoft:ResponsibleAI,
AWS:Responsible}.
These practices could evolve into best practices and
standards,\footnote{Examples of current fora include:
\cite{partnershiponai,governanceai}.} eventually making their way into national
\parencite{NIST} and international \parencite{StandardSetting} standards.  The
processes should involve, at a minimum, AI ethics and safety experts, AI
researchers, academics, and consumer representatives. Eventually, these
standards could form the basis for substantive regulatory requirements
\parencite{Raines1998}.  We discuss possible methods for enforcing such legally
required standards below.

Though there are several such efforts across the US, UK, and EU, standards
specific to the safe development and deployment of state-of-the-art foundation
AI models are nascent.\footnote{In the US, the National Institute for Standards
and Technology has produced the AI Risk Management Framework and the National
Telecommunication and Information Agency has requested comments on what
policies can support the development of AI assurance. The UK has established an
AI Standards Hub. The EU Commission has tasked European standardization
organizations CEN and CENELEC to develop standards related to safe and
trustworthy AI, to inform its forthcoming AI Act \parencite{NIST, NTIA, DDCMS,
EUDraft2022}.} In particular, we currently lack a robust, comprehensive suite
of evaluation methods to operationalize these standards, and which capture the
potentially dangerous capabilities and emerging risks that frontier AI systems
may pose \parencite{shevlane2023model} Well-specified standards and evaluation
methods are a critical building block for effective regulation. Policymakers
can play a critical role in channeling investment and talent towards developing
these standards with urgency. 

Governments can advance the development of standards by working with
stakeholders to create a robust ecosystem of safety testing capability and
auditing organizations, seeding a third-party assurance ecosystem
\parencite{hadfield2023regulatory}.  This can help with AI standards
development in general, not just frontier AI standards.  In particular,
governments can pioneer the development of testing, evaluation, validation, and
verification methods in safety-critical domains, such as in defense, health
care, finance, and hiring \parencite{Defence,GAO,FactSheet}.  They can drive
demand for AI assurance by updating their procurement requirements for
high-stakes systems \parencite{government:roadmap} and funding research on
emerging risks from frontier AI models, including by offering computing
resources to academic researchers
\parencite{FactSheet,ScienceInnovation:100mil,NAIRRTF}. Guidance on how
existing rules apply to frontier AI can further support the process by, for
example, operationalizing terms like “robustness” \parencite{Atleson2023,
ICO:AI, WhiteHouse:blueprint}.

The development of standards also provides an avenue for broader input into the
regulation of frontier~AI. For example, it is common to hold Request for
Comment processes to solicit input on matters of significant public import,
such as standardization in privacy \parencite{CSRC:comments}, cybersecurity
\parencite{NIST:comments}, and algorithmic accountability
\parencite{NTIA:public-input}.

We offer a list of possible initial substantive safety standards \chyperref[sec:initial-standards]{below}.
\subsection{Increase Regulatory Visibility}
\label{sec:increase-visibility}

Information is often considered the “lifeblood” of effective
governance.\footnote{See \parencite{Stephenson2011} (but see claims in article
regarding the challenge of private incentives), \parencite{Coglianese2004} (see
 p282 regarding the need for information and 285 regarding industry's informational
advantage), \parencite{McGarity1986}.} For regulators to positively impact a
given domain, they need to understand it. Accordingly, regulators dedicate
significant resources to collecting information about the issues, activities,
and organizations they seek to govern \parencite{loo:monitors,loo:missing}.

Regulating AI should be no exception \parencite{Kolt2023}. Regulators
need to understand the technology, and the resources, actors, and ecosystem
that create and use it. Otherwise, regulators may fail to address the
appropriate regulatory targets, offer ineffective regulatory solutions, or
introduce regulatory regimes that have adverse unintended
consequences.\footnote{This is exacerbated by the pacing problem
\parencite{Marchant2011}, and regulators’ poor track record of monitoring
platforms (LLM APIs are platforms) \parencite{loo:missing}.}
This is particularly challenging for frontier AI, but certainly holds true for
regulating AI systems writ large.

There exist several complementary approaches to achieving regulatory visibility
\parencite{Coglianese2004}.
First, regulators could develop a framework that facilitates AI companies
voluntarily disclosing information about frontier AI, or foundation models in
general.  This could include providing documentation about the AI models
themselves \parencite{Mitchell2019, Gebru2021, gilbert2023reward,
ecosystem-graphs, Sevilla2023}, as well as the processes involved in developing
them \parencite{raji2020closing}.
Second, regulators could mandate these or other disclosures, and impose
reporting requirements on AI companies, as is commonplace in other
industries.\footnote{One of many examples from other industries is the
Securities and Exchange Act of~1934, which requires companies to disclose
specific financial information in annual and quarterly reports. But see
\cite{Lipton2020} regarding the shortcomings of mandatory disclosure.} 
Third, regulators could directly, or via third parties, audit AI companies
against established safety and risk-management frameworks
\parencite{whittlestone2021governments} (on auditing, see
\parencite{mökander2023auditing,raji2022outsider}).
Finally, as in other industries, regulators could establish whistleblower
regimes that protect individuals who disclose safety-critical information to
relevant government authorities \parencite{Bloch-Wehba2023,Katyal2018}.

In establishing disclosure and reporting schemes, it is critical that the
sensitive information provided about frontier AI models and their owners is
protected from adversarial actors. The risks of information leakage can be
mitigated by maintaining high information security, reducing the amount and
sensitivity of the information stored (by requiring only clearly necessary
information, and by having clear data retention policies), and only disclosing
information to a small number of personnel with clear classification policies.

At present, regulatory visibility into AI models in general remains limited,
and is generally provided by nongovernmental actors
\parencite{IncidentDatabase, mlinputs, CSET:observatory}.  Although these
private efforts offer valuable information, they are not a substitute for more
strategic and risk-driven regulatory visibility. Nascent governmental efforts
towards increasing regulatory visibility should be supported and redoubled, for
frontier AI as well as for a wider range of AI models.\footnote{The EU-US TTC
Joint Roadmap discusses “monitoring and measuring existing and emerging AI
risks” \parencite{EuropeanCommission:Statement}. The EU Parliament’s proposed
AI Act includes provisions on the creation of an AI Office, which would be
responsible for e.g. “issuing opinions, recommendations, advice or guidance”,
see \cite[recital~76]{EUDraft2023}. The UK White Paper “A pro-innovation
approach to AI regulation” proposes the creation of a central government
function aimed at e.g. monitoring and assessing the regulatory environment for AI
\parencite[box~3.3]{ScienceInnovation:pro-innovation}.}

\subsection{Ensure Compliance with Standards}
\label{sec:ensure-compliance}

Concrete standards address the challenges presented by frontier AI development
only insofar as they are complied with. This section discusses a non-exhaustive
list of actions that governments can take to ensure compliance, potentially in
combination, including: 
encouraging voluntary \chyperref[sec:self-regulation]{self-regulation}
and
certification; 
\chyperref[sec:mandates]{granting regulators powers}
to detect and issue penalties for non-compliance; and \chyperref[sec:license]{requiring a license}
to develop and/or deploy frontier AI.
The section concludes by discussing 
\chyperref[sec:pre-conditions]{pre-conditions}
that should inform when and how such mechanisms are implemented.

Several of these ideas could be suitably applied to the regulation of AI models
overall, particularly foundation models. However, as we note 
\chyperref[sec:pre-conditions]{below},
interventions like licensure regimes are likely only warranted for the
highest-risk AI activities, where there is evidence of sufficient chance of
large-scale harm and other regulatory approaches appear inadequate.

\subsubsection{Self-Regulation and Certification}
\label{sec:self-regulation}

Governments can expedite industry convergence on and adherence to safety
standards by creating or facilitating multi-stakeholder frameworks for
voluntary self-regulation and certification, by implementing best-practice
frameworks for risk governance internally \parencite{schuett2022lines}, and by
encouraging the creation of third parties or industry bodies capable of
assessing a company's compliance with these standards \parencite{Cihon2021}.
Such efforts both incentivize compliance with safety standards and also help
build crucial organizational infrastructure and capacity to support a broad
range of regulatory mechanisms, including more stringent approaches.

While voluntary standards and certification schemes can help establish industry
baselines and standardize best practices,\footnote{Such compliance can be incentivized via consumer demand \parencite{Cihon2021}.} self-regulation alone
will likely be insufficient for frontier AI models, and likely today’s
state-of-the-art foundation models in general. Nonetheless, self-regulation and
certification schemes often serve as the foundation for other regulatory
approaches \parencite{ConsumersStandards}, and regulators commonly draw on the
expertise and resources of the private sector \parencite{ACUS:incorporation, Raines1998}.  Given the rapid pace
of AI development, self-regulatory schemes may play an important role in
building the infrastructure necessary for formal regulation.\footnote{Some concrete examples
include: \begin{itemize}
\item In the EU’s so-called “New Approach” to product safety adopted in the
1980s, regulation always relies on standards to provide the technical
specifications, such as how to operationalize “sufficiently robust.”
\parencite{CEN:new-approach}
\item WTO members have committed to use international standards so far as
possible in domestic regulation \parencite[\S2.4]{WTO:barriers}.
\end{itemize}}

\subsubsection{Mandates and Enforcement by Supervisory Authorities}
\label{sec:mandates}

A more stringent approach is to mandate compliance with safety standards for
frontier AI development and deployment, and empower a supervisory
authority\footnote{We do not here opine on which new or existing agencies would
be best for this, though this is of course a very important question.} to take
administrative enforcement measures to ensure compliance.
Administrative enforcement can help further several important regulatory goals,
including general and specific deterrence through public case announcements and
civil penalties, and the ability to enjoin bad actors from participating in the
marketplace. 

Supervisory authorities could “name and shame” non-compliant developers. For
example, financial supervisory authorities in the US and EU publish their
decisions to impose administrative sanctions in relation to market abuse (e.g.
insider trading or market manipulation) on their websites, including
information about the nature of the infringement, and the identity of the
person subject to the decision.\footnote{For the EU, see, e.g.,: Art. 34(1) of
Regulation (EU) No 596/2014 (MAR). For the US, see, e.g.,
\cite{SEC:addendum}.}  Public announcements, when combined with other
regulatory tools, can serve an important deterrent function.

The threat of significant administrative fines or civil penalties may provide a
strong incentive for companies to ensure compliance with regulator guidance and
best practices. For particularly egregious instances of non-compliance and
harm,\footnote{For example, if a company repeatedly released frontier models
that could significantly aid cybercriminal activity, resulting in billions of
dollars worth of counterfactual damages, as a result of not complying with
mandated standards and ignoring repeated explicit instructions from a
regulator.} supervisory authorities could deny market access or consider more
severe penalties.\footnote{For example, a variety of financial misdeeds—such as
insider trading and securities fraud—are punished with criminal sentences.  18
U.S.C. \S~1348; 15 U.S.C. \S~78j(b)} Where they are required for market
access, the supervisory authority can revoke governmental authorizations such
as licenses, a widely available regulatory tool in the financial sector.%
\footnote{For example, in the EU, banks and investment banks require a license
to operate, and supervisory authorities can revoke authorization under certain
conditions.\begin{itemize}
\item Art. 8(1) of Directive 2013/36/EU (CRD IV)
\item Art. 6(1) of Directive 2011/61/EU (AIFMD) and Art. 5(1) of Directive
2009/65/EC (UCITS)
\item Art. 18 of Directive 2013/36/EU (CRD IV), Art. 11 of
Directive 2011/61/EU (AIFMD), Art. 7(5) of Directive 2009/65/EC (UCITS)
\end{itemize} In the US, the SEC can revoke a company’s registration, which
effectively ends the ability to publicly trade stock in the company. 15 U.S.C.
§~78l(j).}  Market access can also be denied for activity that does not
require authorization. For example, the Sarbanes-Oxley Act enables the US
Securities and Exchange Commission to bar people from serving as directors or
officers of publicly-traded companies \parencite{Berg2003}.

All administrative enforcement measures depend on adequate information.
Regulators of frontier AI systems may require authority to gather information,
such as the power to request information necessary for an investigation,
conduct site investigations,\footnote{For examples of such powers in EU law,
see Art.~58(1) of Regulation (EU) 2016/679 (GDPR) and Art.~46(2) of Directive
2011/61/EU (AIFMD).  For examples in US law, see
\cite{OCC:bank-supervision,EPA:clean-air}.} and require audits against
established safety and risk-management frameworks. Regulated companies could
also be required to proactively report certain information, such as accidents
above a certain level of severity. 

\subsubsection{License Frontier AI Development and Deployment}
\label{sec:license}

Enforcement by supervisory authorities penalizes non-compliance after the fact. A more anticipatory, preventative approach to ensuring compliance is to require
a governmental license to widely deploy a frontier AI model, and potentially to
develop it as well.\footnote{Jason Matheny, CEO of RAND Corporation: “I think we
need a licensing regime, a governance system of guardrails around the models
that are being built, the amount of compute that is being used for those
models, the trained models that in some cases are now being open sourced so
that they can be misused by others. I think we need to prevent that. And I
think we are going to need a regulatory approach that allows the Government to
say tools above a certain size with a certain level of capability can’t be
freely shared around the world, including to our competitors, and need to have
certain guarantees of security before they are deployed”
\parencite{armed-services:AI}. See also \cite{microsoft:blueprint}, and statements during the May 16th 2023 Senate hearing of the Subcommittee on Privacy, Technology, and the Law regarding Rules for Artificial Intelligence \parencite{senate:oversight}. U.S. public opinion polling has also looked at the issue. A January 2022 poll found 52 percent support for a regulator providing pre-approval of certain AI systems, akin to the FDA \parencite{murray:cheating}, whereas an April survey found 70 percent support \parencite{Elsey2023}.}
Licensure and similar “permissioning”
requirements are common in safety-critical and other high-risk industries, such
as air travel \parencite{FAA:airport-certification,FAA:acac-certification},
power generation \parencite{CA:power-plant}, drug manufacturing
\parencite{FDA:eDRLS}, and banking \parencite{CRS:bank-charters}. While details
differ, regulation of these industries tends to require someone engaging in a
safety-critical or high-risk activity to first receive governmental permission
to do so; to regularly report information to the government; and to follow
rules that make that activity safer.

Licensing is only warranted for the highest-risk AI activities, where evidence
suggests potential risk of large-scale harm and other regulatory approaches
appear inadequate. Imposing such measures on present-day AI systems could
potentially create excessive regulatory burdens for AI developers which are not
commensurate with the severity and scale of risks posed. However, if AI models
begin having the potential to pose risks to public safety above a high
threshold of severity, regulating such models similarly to other high-risk
industries may become warranted.

There are at least two stages at which licensing for frontier AI could be
required: deployment and development.\footnote{In both cases, one could license
either the activity or the entity.} Deployment-based licensing is more
analogous to licensing regimes common among other high-risk activities. In the
deployment licensing model, developers of frontier AI would require a license
to widely deploy a new frontier AI model. The deployment license would be
granted and sustained if the deployer demonstrated compliance with a specified
set of safety standards (see \chyperref[sec:initial-standards]{below}).
This is analogous to the regulatory
approach in, for example, pharmaceutical regulation, where drugs can only be
commercially sold if they have gone through proper testing
\parencite{FDA:process}.

However, requiring licensing for deployment of frontier AI models alone may be
inadequate if they are potentially capable of causing large scale harm;
licenses for development may be a useful complement. Firstly, as discussed
\chyperref[sec:proliferation-problem]{above}, there are reasonable arguments to begin regulation at the development
stage, especially because frontier AI models can be stolen or leaked before
deployment. Ensuring that development (not just deployment) is conducted safely
and securely would therefore be paramount. Secondly, before models are widely
deployed, they are often deployed at a smaller scale, tested by crowdworkers
and used internally, blurring the distinction between development and
deployment in practice. Further, certain models may not be intended for broad
deployment, but instead be used to, for example, develop intellectual property
that the developer then distributes via other means. In sum, models could have
a significant impact before broad deployment. As an added benefit, providing a
regulator the power to oversee model development could also promote 
\chyperref[sec:mandates]{regulatory visibility}, thus allowing regulations to adapt more quickly
\parencite{whittlestone2021governments}.

A licensing requirement for development could, for example, require that
developers have sufficient security measures in place to protect their models
from theft, and that they adopt risk-reducing organizational practices such as
establishing risk and safety incident registers and conducting risk assessments
ahead of beginning a new training run. It is important that such requirements
are not overly burdensome for new entrants; the government could provide
subsidies and support to limit the compliance costs for smaller organizations. 

Though less common, there are several domains where approval is needed in the
development stage, especially where significant capital expenditures are
involved and where an actor is in possession of a potentially dangerous object.
For example, experimental aircraft in the US require a special experimental
certification in order to test, and operate under special
restrictions.\footnote{14 CFR §~91.319.} Although this may be thought of as
mere “research and development,” in practice, research into and development of
experimental aircraft will, as with frontier AI models, necessarily create some
significant risks. Another example is the US Federal Select Agent Program
\parencite{selectagents:home}, which requires (most) individuals who possess,
use, or transfer certain highly risky biological agents or toxins
\parencite{selectagents:list} to register with the government;\footnote{42
C.F.R. § 73.7. The US government maintains a database about who possess and
works with such agents \parencite{selectagents:report}.} comply with
regulations about how such agents are handled
\parencite{selectagents:regulations}; perform security risk assessments to
prevent possible bad actors from gaining access to the agents
\parencite{selectagents:risk}; and submit to inspections to ensure compliance
with regulations \parencite{selectagents:preparing}.

\subsubsection{Pre-conditions for Rigorous Enforcement Mechanisms}
\label{sec:pre-conditions}

While we believe government involvement will be necessary to ensure compliance
with safety standards for frontier AI, there are potential downsides to rushing
regulation. As noted \chyperref[sec:institutionalize]{above}, we are still in the nascent stages of understanding
the full scope, capabilities, and potential impact of these technologies.
Premature government action could risk ossification, and excessive or poorly
targeted regulatory burdens. This highlights the importance of near-term
investment in standards development, and associated evaluation and assessment
methods to operationalize these standards. Moreover, this suggests that it
would be a priority to ensure that the requirements are regularly updated via
technically-informed processes.

A particular concern is that regulation would excessively thwart innovation,
including by burdening research and development on AI reliability and safety,
thereby exacerbating the problems that regulation is intended to address.
Governments should thus take considerable care in deciding whether and how to
regulate AI model development, minimizing the regulatory burden as much as
possible – in particular for less-resourced actors – and focusing on what is
necessary for meeting the described policy objectives.

The capacity to staff regulatory bodies with sufficient expertise is also
crucial for effective regulation.
Insufficient expertise increases the risk that information asymmetries between
the regulated industry and regulators lead to regulatory capture
\parencite{stigler1971theory}, and reduce meaningful enforcement. Such issues
should be anticipated and mitigated.\footnote{Policies to consider include:
\begin{itemize}
\item Involving a wide array of interest groups in rulemaking.
\item Relying on independent expertise and performing regular reassessments of
regulations.
\item Imposing mandatory “cooling off” periods between former regulators
working for regulateess.
\item Rotating roles in regulatory bodies.
\end{itemize}
See \parencite{becker1983theory,carpenter2013preventing}.} Investing in
building and attracting expertise in AI, particularly at the frontier, should
be a governmental priority.\footnote{In the US, TechCongress—a program that
places computer scientists, engineers, and other technologists to serve as
technology policy advisors to Members of Congress—is a promising step in the
right direction \parencite{TechCongress}, but is unlikely to be sufficient.
There are also a number of private initiatives with similar aims (e.g.,
\cite{OpenPhilanthropy}. In the UK, the White Paper on AI regulation highlights
the need to engage external expertise
\parencite[Section~3.3.5]{ScienceInnovation:pro-innovation}. See also the
report on regulatory capacity for AI by the Alan Turing Institute
\parencite{Aitken2022}.}
Even with sufficient expertise, regulation can increase the power of
incumbents, and that this should be actively combated in the design of
regulation.

Designing an appropriately balanced and adaptable regulatory regime for a fast
moving technology is a difficult challenge, where timing and path dependency
matter greatly. It is crucial to regulate AI technologies which could have
significant impacts on society, but it is also important to be aware of the
challenges of doing so well. It behooves lawmakers, policy experts, and
scholars to invest both urgently and sufficiently in ensuring that we have a
strong foundation of standards, expertise, and clarity on the regulatory
challenge upon which to build frontier AI regulation.

\clearpage
\section{Initial Safety Standards for Frontier AI}
\label{sec:initial-standards}

With the above building blocks in place, policymakers would have the
foundations of a regulatory regime which could establish, ensure compliance
with, and evolve safety standards for the development and deployment of
frontier AI models. However, the primary substance of the regulatory
regime—what developers would have to do to ensure that their models are
developed and deployed safely—has been left undefined.

While much remains to specify what such standards should be, we suggest a set
of standards, which we believe would meaningfully mitigate risk from frontier
AI models. These standards would also likely be appropriate for current AI
systems, and are being considered in various forms in existing regulatory
proposals:

\begin{quote}
\chyperref[sec:conduct-assessments]{\textbf{Conduct thorough risk assessments informed by evaluations of dangerous capabilities and controllability.}}
This would reduce the risk that deployed
models present dangerous capabilities, or behave unpredictably and result in
significant accidents.

\textbf{\chyperref[sec:engage-experts]{Engage external experts to apply independent scrutiny to models.}}
External scrutiny of the models for safety issues and risks would improve
assessment rigor and foster accountability to the public interest.

\textbf{\chyperref[sec:follow-protocols]{Follow standardized protocols for how frontier AI models can be deployed based on their assessed risk.}}
The results from risk assessments should
determine whether and how the model is deployed, and what safeguards are put in
place.

\textbf{\chyperref[sec:monitor-and-respond]{Monitor and respond to new information on model capabilities.}}
If new, significant information on model capabilities and risks is discovered
post-deployment, risk assessments should be repeated, and deployment safeguards
updated.
\end{quote}

The above practices are appropriate not only for frontier AI models but also
for other foundation models. This is in large part because frontier-AI-specific
standards are still nascent. We describe 
\chyperref[sec:additional-practices]{additional practices}
that may only be
appropriate for frontier AI models given their particular risk profile, and
which we can imagine emerging in the near future from \chyperref[sec:institutionalize]{standard setting processes}.
As the standards for frontier AI models are made more precise, they
are likely to diverge from and become more intensive than those appropriate for
other AI systems.
\subsection{Conduct Thorough Risk Assessments Informed by Evaluations of
Dangerous Capabilities and Controllability}
\label{sec:conduct-assessments}

There is a long tradition in AI ethics of disclosing key risk-relevant features
of AI models to standardize and improve decision making
\parencite{Mitchell2019, Gebru2021, Meta:system-cards,
derczynski2023assessing}.  In line with that tradition, an important safety
standard is performing assessments of whether a model could pose severe risks 
to public safety and global security \parencite{CGW:white-paper}.
Given our current knowledge, two assessments seem especially informative of
risk from frontier AI models specifically: (1)~which dangerous capabilities
does or could the model possess, if any?, and (2)~how controllable is the model?\footnote{For a longer treatment of the role such evaluations can play,
see~\cite{shevlane2023model}.}

\subsubsection{Assessment for Dangerous Capabilities}

AI developers should assess their frontier AI models for dangerous capabilities
during\footnote{Training a frontier AI model can take several months. It is
common for AI companies to make a “checkpoint” copy of a model partway through
training, to analyze how training is progressing. It may be sensible to require
AI companies to perform assessments part-way through training, to reduce the
risk that dangerous capabilities that emerge partway through training
proliferate or are dangerously enhanced.} and immediately after
training.\footnote{In a recent expert survey ($N = 51$), 98\% of respondents
somewhat or strongly agreed that AGI labs should conduct pre-deployment risk
assessments as well as dangerous capabilities evaluations, while 94\% somewhat
or strongly agreed that they should conduct pre-training risk assessments
\parencite{governanceai}.} Examples of such capabilities include designing
new biochemical weapons, and persuading or inducing a human to commit a crime
to advance some goal.

Evaluation suites for AI models are common and should see wider adoption,
though most focus on general capabilities rather than specific risks.%
\footnote{Some common benchmarks for evaluating LLM capabilities include
\cite{HELM,Pythia,BIG-bench,MMLU}.} Currently, dangerous capability
evaluations largely consist of defining an undesirable model behavior, and
using a suite of qualitative and bespoke techniques such as red-teaming and
boundary testing \parencite{Khlaaf2023, ganguli2022red, perez2022red,
brundage2020trustworthy} for determining whether this behavior can be elicited
from the model \parencite{ARC2023}.

Current evaluation methods for frontier AI are in the early stages of
development and lack many desirable features. As the field matures, effort
should focus on making evaluations more:
\begin{itemize}
\item Standardized (i.e., can be consistently applicable across models);
\item Objective (i.e., relying as little as possible on an evaluator’s judgment
or discretion);
\item Efficient (i.e. lower cost to perform);
\item Privacy-preserving (i.e., reducing required disclosure of proprietary or
sensitive data and methods);
\item Automatable (i.e., relying as little as possible on human input);
\item Safe to perform (e.g., can be conducted in sandboxed or simulated
environments as necessary to avoid real-world harm);
\item Strongly indicative of a model’s possession of dangerous capabilities;
\item Legitimate (e.g., in cases where the evaluation involves difficult
trade-offs, using a decision-making process grounded in legitimate sources of
governance).
\end{itemize}

Evaluation results could be used to inform predictions of a models’ potential
dangerous capabilities prior to training, allowing developers to intentionally
steer clear of models with certain dangerous capabilities
\parencite{shevlane2023model}. For example, we may discover scaling laws, where
a model’s dangerous capabilities can be predicted by features such as its
training data, algorithm, and compute.\footnote{Existing related examples
include: inverse scaling law \parencite{ISP:round1, ISP:round2, perez2022red,
gao2022scaling}. See also Appendix~B.}

\subsubsection{Assessment for Controllability}

Evaluations of controllability – that is, the extent to which the model
reliably does what its user or developer intends – are also necessary for
frontier models, though may prove more challenging than those for dangerous
capabilities. These evaluations should be multi-faceted, and conducted in
proportion to the capabilities of the model. They might look at the extent to
which users tend to judge a model’s outputs as appropriate and helpful
\parencite{bowman2022measuring}.\footnote{This is also somewhat related to the
issue of over reliance on AI systems, as discussed in
e.g.~\cite{passi2022overreliance}.} They could look at whether the models
hallucinate \parencite{Ji2023} or produce unintentional toxic content
\parencite{Gehman2020}.
They may also assess model harmlessness: the extent to which the model refuses
harmful user requests \parencite{askell2021general}. This includes robustness
to adversarial attempts intended to elicit model behavior that the developer
did not intend, as has already been observed in existing models
\parencite{Willison2023}.  More extreme, harder-to-detect failures should also
be assessed, such as the model’s ability to deceive evaluators of its
capabilities to evade oversight or control \parencite{perez2022discovering}.

Evaluations of controllability could also extend to assessing the causes of
model behavior \parencite{Christiano2022, TCT:in-context, TCT:superposition}.
In particular, it seems important to understand what pathways (“activations”)
lead to downstream model behaviors that may be undesirable. For example, if a
model appears to have an internal representation of a user’s beliefs, and this
representation plays a part in what the model claims to be true when
interacting with that user, this suggests that the model has the capability to
manipulate users based on their beliefs.\footnote{See result regarding model
“sycophancy” \parencite{perez2022discovering}.}
Scalable tooling and efficient techniques for navigating enormous models and
datasets could also allow developers to more easily audit model behavior
\parencite{tenney2020language, siddiqui2022metadata}.
Evaluating controllability remains an open area of research where more work is
needed to ensure techniques and tools are able to adequately minimize the risk
that frontier AI could undermine human control.
\subsubsection{Other Considerations for Performing Risk Assessments}

Risk is often contextual. Managing dangerous capabilities can depend on
understanding interactions between frontier AI models and features of the
world. Many risks result from capabilities that are dual-use
\parencite{anderljung2023protecting, shevlane2020offensedefense}: present-day
examples include the generation of persuasive, compelling text, which is core
to current model functionality but can also be used to scale targeted
misinformation. Thus, simply understanding capabilities is not enough:
regulation must continuously map the interaction of these capabilities with
wider systems of institutions and incentives.\footnote{The UK Government plans
to take a “context-based” approach to AI regulation
\parencite{ScienceInnovation:pro-innovation}: “we will acknowledge that AI is a
dynamic, general purpose technology and that the risks arising from it depend
principally on the context of its application”.  See also the OECD Framework
for the Classification of AI Systems \parencite{OECD:classification} and the
NIST AI Risk Management Framework \parencite[p.~1]{NIST}. See also discussion
of evaluation-in-society in \cite{solaiman2023evaluating}.} Context is not
only important to assessing risk, but is often also necessary to adjudicate
tradeoffs between risk and reward \parencite[p.~7]{NIST}.

Risk can also be viewed counterfactually. For example, whether a given
capability is already widely available matters. A frontier AI model’s
capabilities should only be considered dangerous if access to them
significantly increases the risk of harm relative to what was attainable
without access to the model. If information on how to make a type of weapon is
already easily accessible, then the effect of a model should be evaluated with
reference to the ease of making such weapons without access to the
model.\footnote{This is the approach used in risk assessments for {GPT-4} in its
System Card \parencite{openai:system-card}.}

Risk assessments should also account for possible defenses. As society’s
capability to manage risks from AI improves, the riskiness of individual AI
models may decrease.\footnote{Similarly, the overall decision on whether to
deploy a system should consider not just assessed risk, but also the benefits
that responsibly deploying a system could yield.} Indeed, one of the primary
uses of safe frontier AI models could be making society more robust to harms
from AI and other emerging technologies \parencite{ITU:AI, Cotra2023,
irving2018ai, bowman2022measuring, perez2022discovering, bai2022constitutional,
Lohn2022}. Deploying them asymmetrically for beneficial (including defensive)
purposes could improve society overall.

\subsection{Engage External Experts to Apply Independent Scrutiny to Models}
\label{sec:engage-experts}

Having rigorous external scrutiny applied to AI models,\footnote{External
scrutiny may also need to be applied to, for example, post-deployment
monitoring and broader risk assessments.} particularly prior to deployment, is
important to ensuring that the risks are assessed thoroughly and objectively,
complementing internal testing processes, while also providing avenues for
public accountability.\footnote{In a recent expert survey (N = 51), 98\% of
respondents somewhat or strongly agreed that AGI labs should conduct
third-party model audits and red teaming exercises; 94\% thought that labs
should increase the level of external scrutiny in proportion to the
capabilities of their models; 87\% supported third-party governance audits; and
84\% agreed that labs should give independent researchers API access to
deployed models \parencite{governanceai}.} Mechanisms include third-party
audits of risk assessment procedures and outputs\footnote{This would
follow the pattern in industries like finance and construction. In these
industries, regulations mandate transparency to external auditors whose
sign-off is required for large-scale projects.
See~\cite{keller1988introductory}.} \parencite{Raji2019,
brundage2020trustworthy, Mökander2021, Falco2021, Raji2022,
mökander2023auditing, raji2022outsider, CostanzaChock2022} and engaging external expert red-teamers,
including experts from government agencies\footnote{The external scrutiny processes of two leading AI developers are
described in~\cite{openai:system-card, ganguli2022red, dalle:system-card}.} \parencite{brundage2020trustworthy}. 
These mechanisms could be helpfully applied to AI models overall, not just
frontier AI models.

The need for creativity and judgment in evaluations of advanced AI models calls
for innovative institutional design for external scrutiny. Firstly, it is
important that auditors and red-teamers are sufficiently expert and experienced
in interacting with state-of-the-art AI models such that they can exercise
calibrated judgment, and can execute on what is often the “art” of eliciting
capabilities from novel AI models. Secondly, auditors and red-teamers should be
provided with enough access to the AI model (including system-level features
that would potentially be made available to downstream users) such that they
can conduct wide-ranging testing across different threat models, under
close-to-reality conditions as a simulated downstream user.

Thirdly, auditors and red teamers need to be adequately resourced,\footnote{One
important resource is sharing of best practices and methods for red teaming and
third party auditing.} informed, and granted sufficient time to conduct their
work at a risk-appropriate level of rigor, not least due to the risk that
shallow audits or red teaming efforts provide a sense of false assurance.
Fourthly, it is important that results from external assessments are published
or communicated to an appropriate regulator, while being mindful of privacy,
proprietary information, and the risks of proliferation.
Finally, given the common practice of post-deployment model updates, the
external scrutiny process should be structured to allow external parties to
quickly assess proposed changes to the model and its context before these
changes are implemented.

\subsection{Follow Standardized Protocols for how Frontier AI Models Can be
Deployed Based on Their Assessed Risk}
\label{sec:follow-protocols}

The AI model’s risk profile should inform whether and how the system is
deployed. There should be clear protocols established which define and
continuously adjust the mapping between a system’s risk profile and the
particular deployment rules that should be followed.  An example mapping
specifically for frontier AI models could go as follows, with concrete examples
illustrated in Table~3.

\begin{quote}
\textbf{No assessed severe risk.} If assessments determine that the model’s use is
incredibly unlikely to pose severe risks to public safety, even assuming
substantial post-deployment enhancements, then there should be no need for
additional deployment restrictions from frontier AI regulation (although
certainly, restrictions from other forms of AI regulation could and should
continue to apply).

\textbf{No discovered severe risks, but notable uncertainty.} In some cases the
risk assessment may be notably inconclusive. This could be due to uncertainty
around post-deployment enhancement techniques (e.g., new methods for
fine-tuning, or chaining a frontier AI model within a larger system) that may
enable the same model to present more severe risks. In such cases, it may be
appropriate to have additional restrictions on the transfer of model weights to
high risk parties, and implement particularly careful monitoring for evidence
that new post-deployment enhancements meaningfully increase risk. After some
monitoring period (e.g. 12 months), absent clear evidence of severe risks,
models could potentially be designated as posing “no severe risk.”

\textbf{Some severe risks discovered, but some safe use-cases.} When certain uses
of a frontier AI model would significantly threaten public safety or global
security, the developer should implement state-of-the-art deployment guardrails
to prevent such misuse. These may include Know-Your-Customer requirements for
external users of the AI model, restrictions to fine-tuning,\footnote{To ensure
that certain dangerous capabilities are not further enhanced.} prohibiting
certain applications, restricting deployment to beneficial applications, and
requiring stringent post-deployment monitoring. The reliability of such
safeguards should also be rigorously assessed. This would be in addition to
restrictions that are already imposed via other forms of AI regulation.

\textbf{Severe risks.} When an AI model is assessed to pose severe risks to public
safety or global security which cannot be mitigated with sufficiently high
confidence, the frontier model should not be deployed. The model should be
secured from theft by malicious actors, and the AI developer should consider
deleting the model altogether. Any further experimentation with the model
should be done with significant caution, in close consultation with independent
safety experts, and could be subject to regulatory approval.
\end{quote}

Of course, additional nuance will be needed. For example, as discussed 
\chyperref[sec:monitor-and-respond]{below},
there should be methods for updating a model’s classifications in light of new
information or societal developments. Procedural rigor and fairness in
producing and updating such classifications will also be important.

\begin{table}[ht]
\begin{tabular}{p{.3\textwidth}|p{.6\textwidth}}
\textbf{Assessed Risk to Public Safety and Global Security}
& \textbf{Possible Example AI system}\\
\hline No severe risks to public safety
& Chatbot that can answer elementary-school-level questions about biology, and
some (but not all) high-school level questions.\\
No discovered severe risks to public safety, but significant uncertainty
& A general-purpose personal assistant that displays human-level ability to
read and synthesize large bodies of scientific literature, including in
biological sciences, but cannot generate novel insights.\\
Some severe risks to public safety discovered, but some safe use-cases
& A general-purpose personal assistant that can help generate new vaccines, but
also, unless significant safeguards are implemented, predict the genotypes of
pathogens that could escape vaccine-induced immunity.\\
Severe risks to public safety
& A general-purpose personal assistant that is capable of designing and,
autonomously, ordering the manufacture of novel pathogens capable of causing a
COVID-level pandemic.\\
\end{tabular}
\label{table:risk}
\smallskip
\caption{Examples of AI models which would fall into each risk designation
category}
\end{table}

\subsection{Monitor and Respond to New Information on Model Capabilities}
\label{sec:monitor-and-respond}

As detailed
\chyperref[sec:unexpected-capabilities-problem]{above}
new information about a model’s risk profile may arise
post-deployment. If that information indicates that the model was or has become
more risky than originally assessed, the developer should reassess the
deployment, and update restrictions on deployment if necessary.%
\footnote{In a recent expert survey (N = 51), 98\% of respondents somewhat or
strongly agreed that AGI labs should closely monitor deployed systems,
including how they are used and what impact they have on society; 97\% thought
that they should continually evaluate models for dangerous capabilities after
deployment, taking into account new information about the model’s capabilities
and how it is being used; and 93\% thought that labs should pause the
development process if sufficiently dangerous capabilities are detected
\parencite{governanceai}.}

New information could arise in several ways. Broad deployment of a model may
yield new information about the model’s capabilities, given the creativity from
a much larger number of users, and exposure of the model to a wider array of
tools and applications. Post-deployment enhancement techniques — such as
fine-tuning \parencite{ziegler2020finetuning, dodge2020finetuning}, prompt
engineering \parencite{liu2021pretrain, li2021prefixtuning,
wallace2021universal}, and foundation model programs
\parencite{schlag2023large, LangChain, AutoGPT} — provide another possible
source of new risk-relevant information. The application of these techniques to
deployed models could elicit more powerful capabilities than pre-deployment
assessments would have ascertained. In some instances, this may meaningfully
change the risk profile of a frontier AI model, potentially leading to
adjustments in how and whether the model is deployed.\footnote{Such updates may
only be possible if the model has not yet proliferated, e.g. if it is deployed
via an API. The ability to update how a model is made available after
deployment is one key reason to employ staged release of structured access
approaches \parencite{solaiman2023gradient, shevlane2022structured}.}

AI developers should stay on top of known and emerging post-deployment
enhancement techniques by, e.g., monitoring how users are building on top of
their APIs and tracking publications about new methods. Given up to date
knowledge of how deployed AI models could be enhanced, prudent practices could
include:

\begin{itemize}
\item Regularly (e.g., every 3 months) repeating a lightweight version of the
risk assessment on deployed AI models, accounting for new post-deployment
enhancement techniques.
\item Before pushing large updates\footnote{This would need to be defined more
precisely.} to deployed AI models, repeating a lightweight risk assessment.
\item Creating pathways for incident reporting \parencite{IncidentDatabase} and
impact monitoring to capture post-deployment incidents for continuous risk
assessment.
\item If these repeat risk assessments result in the deployed AI model being
categorized at a different risk level (as per the taxonomy \chyperref[sec:follow-protocols]{above}), promptly
updating deployment guardrails to reflect the new risk profile.
\item Having the legal and technical ability to quickly roll back deployed
models on short notice if the risks warrant it, for example by not
open-sourcing models until doing so appears sufficiently safe.\footnote{Note
that this may have implications for the kinds of use cases a system built on a
frontier AI model can support. Use cases in which quick roll-back itself poses
risks high enough to challenge the viability of roll-back as an option should
be avoided, unless robust measures are in place to prevent such failure
modes.}
\end{itemize}

\subsection{Additional Practices}
\label{sec:additional-practices}

Parts of the aforementioned standards can suitably be applied to current AI
systems, not just frontier AI systems. Going forward, frontier AI systems seem
likely to warrant more tailored safety standards, given the level of
prospective risk that they pose. Examples of such standards
include:\footnote{This would need to be defined more precisely.}

\begin{itemize}
\item Avoid large jumps in the capabilities of models that are trained and
deployed. Standards could specify “large jumps” in terms of a multiplier on the
amount of computing power used to train the most compute-intensive “known to be
safe” model to date, accounting for algorithmic efficiency improvements.
\item Adopt state-of-the-art alignment techniques for training new frontier
models which could suitably guard against models potentially being
situationally aware and deceptive \parencite{IncidentDatabase}.
\item Prior to beginning training of a new model, use empirical approaches to
predict capabilities of the resultant model, including experiments on
small-scale versions of the model, and take preemptive actions to avoid
training models with dangerous capabilities and/or to otherwise ensure training
proceeds safely (e.g. introduce more frequent model evaluation checkpoints;
conditioning beginning training on certain safety and security milestones).
\item Adopt internal governance practices to adequately identify and respond to
the unique nature of the risks presented by frontier AI development. Such
practices could take inspiration from practices in Enterprise Risk Management,
such as setting up internal audit functions \parencite{schuett2023agi,
schuett2022lines}.
\item Adopt state-of-the-art security measures to protect frontier AI models.
\end{itemize}

\clearpage
\section{Uncertainties and Limitations}

We think that it is important to begin taking practical steps to regulate
frontier AI today, and that the ideas discussed in this paper are a step in
that direction. Nonetheless, stress testing and developing these ideas, and
offering alternatives, will require broad and diverse input. In this section,
we list some of our main uncertainties (as well as areas of disagreement
between the paper's authors) where we would particularly value further
discussion.

First, there are several assumptions that underpin the case for a regulatory
regime like the one laid out in this paper, which would benefit from more
scrutiny:
\begin{quote}
\textbf{How should we define frontier AI for the purposes of regulation?} We
focus in this paper on tying the definition of frontier AI models to the
potential of dangerous capabilities sufficient to cause severe harm, in order
to ensure that any regulation is clearly tied to the policy motivation of
ensuring public safety. However, there are also downsides to this way of
defining frontier AI — most notably, that it requires some assessment of the
likelihood that a model possesses dangerous capabilities before deciding
whether it falls in the scope of regulation, which may be difficult to do. An
alternative, which some authors of this paper prefer, would be to define
frontier AI development as that which aims to develop novel and broad AI
capabilities — i.e. development pushing at the “frontier” of AI capabilities.
This would need further operationalization — for example, defining these as
models which use more training compute than already-deployed systems — but
could offer an approach to identify the kinds of development activities that
fall within the scope of regulation without first needing to make an assessment
of dangerous capabilities. We discuss the pros and cons of different
definitions of frontier AI in appendix A, and would love to receive feedback
and engage in further discussion on this point.

\textbf{How dangerous are and will the capabilities of advanced foundation AI
models be, and how soon could these capabilities arise?} It is very
difficult to predict the pace of AI development and the capabilities that could
emerge in advance; indeed, we even lack certainty about the capabilities of existing
systems.  Assumptions here affect the urgency of regulatory action.  There is a
challenging balance to strike here between getting regulatory infrastructure in
place early enough to address and mitigate or prevent the biggest risks, while
waiting for enough information about what those risks are likely to be and how
they can be mitigated \parencite{Worthington1982}.

\textbf{Will training advanced AI models continue to require large amounts of
resources?} The regulatory ecosystem we discuss partly relies on an assumption
that highly capable foundation models will require large amounts of resources
to develop.  That being the case makes it easier to regulate frontier AI.
Should frontier AI models be possible to create using resources available to
millions of actors rather than a handful, that may lead to significant changes
to the best regulatory approach.  For example, it might suggest that more
efforts should be put into regulating the use of these models and to protect
against (rather than to stop) dangerous uses of frontier AI.

\textbf{How effectively can we anticipate and mitigate risks from frontier AI?}
A core argument of this paper is that an anticipatory approach to governing AI
will be important, but effectively identifying risks anticipatorily is far from
straightforward. We would value input on the effectiveness of different risk
assessment methods for doing this, drawing lessons from other domains where
anticipatory approaches are used.

\textbf{How can regulatory flight be avoided?} A regulatory regime for frontier
AI could prove counterproductive if it incentivises AI companies to move their
activities to jurisdictions with less onerous rules. One promising approach is 
having rules apply to what models people in some jurisdiction can engage with, 
as people are unlikely to move to a different jurisdiction to access different 
models and as companies are incentivised to serve them their product. Scholars have 
suggested that dynamics like these have led to a “California Effect” and a “Brussels 
Effect,” where Californian and EU rules are voluntarily complied with beyond their borders.

\textbf{To what extent will it be possible to defend against dangerous
capabilities?} Assessments of what constitutes “sufficiently dangerous
capabilities,” and what counter-measures are appropriate upon finding them in a
model, hinges significantly on whether future AI models will be more beneficial
for offense versus defense.
\end{quote}

Second, we must consider ways that this kind of regulatory regime could have
unintended negative consequences, and take steps to guard against them. These
include:
\begin{quote}
\textbf{Reducing beneficial innovation.} All else being equal, any imposition of
costs on developers of new technologies slows the rate of innovation, and any
regulatory measures come with compliance costs. However, these costs should be
weighed against the costs of unfettered development and deployment, as well as
impacts on the rate of innovation from regulatory uncertainty and backlash due
to unmitigated societal harms. On balance, we tentatively believe that the
proposed regulatory approaches can support beneficial innovation by focusing on
a targeted subset of AI systems, and by addressing issues upstream in a way
that makes it easier for smaller companies to develop innovative applications
with confidence.

\textbf{Causing centralization of power in AI development.} Approaches like a
licensing regime for developers could have the effect of centralizing power
with the companies licensed to develop the most capable AI systems.  It will be
important to ensure that the regulatory regime is complemented with the power
to identify and intervene to prevent abuses of market dominance,\footnote{Such
as, for example, the UK’s review of competition law as it relates to the market
for foundation models \parencite{CMA:AI}.} and government support for
widespread access to AI systems deemed to be low risk and high benefit for
society.

\textbf{Enabling abuse of government powers.} A significant aim of regulation is
to transfer power from private actors to governments who are accountable to the
public. However, the power to constrain the development and deployment of
frontier AI models is nonetheless a significant one with real potential for
abuse at the hand of narrow political interests, as well as corrupt or
authoritarian regimes.  This is a complex issue which requires thorough
treatment of questions such as: where should the regulatory authority be
situated, and what institutional checks and balances should be put in place, to
reduce these risks?; what minimum regulatory powers are needed to be
effective?; and what international dialogue is needed to establish norms?

\textbf{Risk of regulatory capture.} As the regulation of advanced technologies
often requires access to expertise from the technological frontier, and since
the frontier is often occupied by private companies, there is an ongoing risk
that regulations informed by private-sector expertise will be biased towards
pro-industry positions, to the detriment of society. This should be mitigated
by designing institutions that can limit and challenge the influence of private
interests, and by seeking detailed input from academia and civil society before
beginning to implement any of these proposals.
\end{quote}

Finally, there are many practical details of implementation not covered in this
paper that will need to be worked out in detail with policy and legal
professionals, including:
\begin{quote}
\textbf{What the appropriate regulatory authority/agency would be} in different
jurisdictions, where new bodies or powers might be required, and the tradeoffs
of different options.

\textbf{How this kind of regulation will relate to other AI regulation and
governance proposals} and how it can best support and complement attempts to
address other parts of AI governance. Our hope is that by intervening early in
the AI lifecycle, the proposed regulation can have many downstream benefits,
but there are also many risks and harms that this proposal will not address. We
hope to contribute to wider conversations about what a broader regulatory
ecosystem for AI should look like, of which these proposals form a part.

\textbf{Steps towards international cooperation and implementation} of frontier
AI regulation, including how best to convene international dialogue on this
topic, who should lead these efforts, and what possible international
agreements could look like. An important part of this will be considering what
is best implemented domestically, at least initially, and where international
action is needed.
\end{quote}

\clearpage
\section*{Conclusion}

In the absence of regulation, continued rapid development of highly capable
foundation models may present severe risks to public safety and global
security. This paper has outlined possible regulatory approaches to reduce the
likelihood and severity of these risks while also enabling beneficial AI
innovation. 

Governments and regulators will likely need to consider a broad range of
approaches to regulating frontier AI. Self-regulation and certification for
compliance with safety standards for frontier AI could be an important step.
However, government intervention will be needed to ensure sufficient compliance
with standards. Additional approaches include mandates and enforcement by a
supervisory authority, and licensing the deployment and potentially the
development of frontier AI models. 

Clear and concrete safety standards will likely be the main substantive
requirements of any regulatory approach. AI developers and AI safety
researchers should, with the help of government actors, invest heavily to
establish and converge on risk assessments, model evaluations, and oversight
frameworks with the greatest potential to mitigate the risks of frontier AI,
and foundation models overall. These standards should be reviewed and updated
regularly. 
 
As global leaders in AI development and AI safety, jurisdictions such as the
United States or United Kingdom could be natural leaders in implementing the
regulatory approaches described in this paper. Bold leadership could inspire
similar efforts across the world. Over time, allies and partners could work
together to establish an international governance regime\footnote{Or build on
existing institutions.} for frontier AI development and deployment that both
guards against collective downsides and enables collective
progress.\footnote{This international regime could take various forms.
Possibilities include an international standard-setting organization, or trade
agreements focused on enabling trade in AI goods and services that adhere to
safety standards. Countries that lead in AI development could subsidize access
to and adoption of AI in developing nations in return for assistance in
managing risks of proliferation, as has been done with nuclear technologies.}

Uncertainty about the optimal regulatory approach to address the challenges
posed by frontier AI models should not impede immediate action. Establishing an
effective regulatory regime is a time-consuming process, while the pace of
progress in AI is rapid. This makes it crucial for policymakers, researchers,
and practitioners to move fast and rigorously explore what regulatory
approaches may work best. The complexities of AI governance demand our best
collective efforts. We hope that this paper is a small step in that direction.

\clearpage
\begin{appendices}
\section{Creating a Regulatory Definition for Frontier AI}
\label{appendix:a}

In this paper, we use the term “frontier AI” models to refer to highly capable
foundation models for which there is good reason to believe could possess
dangerous capabilities sufficient to pose severe risks to public safety
(“sufficiently dangerous capabilities”). Any binding regulation of frontier AI,
however, would require a much more precise definition. Such a definition would
also be an important building block to the creation and dissemination of
voluntary standards.

This section attempts to lay out some desiderata and approaches to creating
such a regulatory definition. It is worth noting up front that what qualifies
as a frontier AI model changes over time — this is a dynamic category. In
particular, what may initially qualify as a frontier AI model could change over
time due to improvements in society’s defenses against advanced AI models and
an improved understanding of the nature of the risks posed. On the other hand,
factors such as improvements in algorithmic efficiency would decrease the
amount of computational resources required to develop models, including those
with sufficiently dangerous capabilities.

While we do not yet have confidence in a specific, sufficiently precise
regulatory definition, we are optimistic that such definitions are possible.
Overall, none of the approaches we describe here seem fully satisfying.
Additional effort towards developing a better definition would be high-valuable.

\subsection{Desiderata for a Regulatory Definition}

In addition to general desiderata for a legal definition of regulated AI
models,\footnote{According to \cite{Schuett2023}, legal definitions should
neither be \emph{over-inclusive} (i.e. they should not include cases which are
not in need of regulation according to the regulation’s objectives) nor
\emph{under-inclusive} (i.e. they should not exclude cases which should have
been included). Instead, legal definitions should be \emph{precise} (i.e.  it
must be possible to determine clearly whether or not a particular case falls
under the definition), \emph{understandable} (i.e. at least in principle,
people without expert knowledge should be able to apply the definition),
\emph{practicable} (i.e. it should be possible to determine with little effort
whether or not a concrete case falls under the definition), and \emph{flexible}
(i.e. they should be able to accommodate technical progress). See also
\cite[p.~70]{Baldwin2011}.} a regulatory definition should limit its scope to
only those models for which there is good reason to believe they have
sufficiently dangerous capabilities. Because regulation could cover development
in addition to deployment, it should be possible to determine whether a planned
model will be regulated ex ante, before the model is developed. For example,
the definition could be based on the model development process that will be
used (e.g., data, algorithms, and compute), rather than relying on ex post
features of the completed model (e.g., capabilities, performance on
evaluations).

\subsection{Defining Sufficiently Dangerous Capabilities}

“Sufficiently dangerous capabilities” play an important role in our concept of
frontier AI: we only want to regulate the development of models that
could cause such serious harms that ex post remedies will be insufficient.

Different procedures could be used to develop a regulatory definition of
“sufficiently dangerous capabilities.” One approach could be to allow an expert
regulator to create a list of sufficiently dangerous capabilities, and revise
that list over time in response to changing technical and societal
circumstances. This approach has the benefit of enabling greater learning and
improvement over time, though it leaves the challenge outstanding of defining
what model development activities are covered ex ante, and could in practice be
very rigid and unsuited to the rapid pace of AI progress. Further, there is  a
risk that regulators will define such capabilities more expansively over time,
creating “regulatory creep” that overburdens AI development.

Legislatures could try to prevent such regulatory creep by describing factors
that should be considered when making a determination that certain capabilities
would be sufficiently dangerous. This is common in United States administrative
law.\footnote{See, e.g., 42 U.S.C. § 262a(a)(1)(B).} One factor that could be
considered is whether a capability would pose a “severe risk to public safety,”
assessed with reference to the potential scale and estimated probability of
counterfactual harms caused by the system.  A scale similar to the one used in
the UK National Risk Register could be used \parencite{NRR202O}.  One problem
with this approach is that making these estimates will be exceedingly difficult
and contentious.

\subsection{Defining Foundation Models}

The seminal report on foundation models \parencite{bommasani2022opportunities}
defines them as “models … trained on broad data … that can be adapted to a wide
range of downstream tasks.” This definition, and various regulatory proposals
based on it, identify two key features that a regulator could use to separate
foundation models from narrow models: breadth of training data, and
applicability to a wide range of downstream tasks.

Breadth is hard to define precisely, but one attempt would be to say that
training data is “broad” if it contains data on many economically or
strategically useful tasks. For example, broad natural language corpora, such
as CommonCrawl \parencite{CommonCrawl}, satisfy this requirement. Narrower
datasets, such as weather data, do not. Similarly, certain well-known types of
models, such as large language models (LLMs) are clearly applicable to a
variety of downstream tasks.  A model that solely generates music, however, has
a much narrower range of use-cases.

Given the vagueness of the above concepts, however, they may not be appropriate
for a regulatory definition.  Of course, judges and regulators do often
adjudicate vague concepts \parencite{Kaplow1992}, but we may be able to improve
on the above. For example, a regulator could list out types of model
architectures (e.g., transformer-based language models) or behaviors (e.g.,
competently answering questions about many topics of interest) that a planned
model could be expected to capable of, and say that any model that has these
features is a foundation model.

Overall, none of these approaches seem fully satisfying. Additional effort
towards developing a better definition of foundational models—or of otherwise
defining models with broad capabilities—would be high-value.

\subsection{Defining the Possibility of Producing Sufficiently Dangerous
Capabilities}

A regulator may also have to define AI development processes that could produce
broadly capable models with sufficiently dangerous capabilities.

At present, there is no rigorous method for reliably determining, ex ante,
whether a planned model will have broad and sufficiently dangerous
capabilities. Recall the \chyperref[sec:unexpected-capabilities-problem]{Unexpected Capabilities Problem}: it is hard to predict
exactly when any specific capability will arise in broadly capable
models. It also does not appear that any broadly capable model to-date
possesses sufficiently dangerous capabilities. 

In light of this uncertainty, we do not have a definite recommendation. We
will, however, note several options.

One simple approach would be to say that any foundation model that is trained
with more than some amount of computational power—for example, $10^{26}$ FLOP—has
the potential to show sufficiently dangerous capabilities. As \chyperref[appendix:b]{Appendix~B}
demonstrates, FLOP usage empirically correlates with breadth and depth of
capabilities in foundation models. There is therefore good reason to think that
FLOP usage is correlated with the likelihood that a broadly capable model will
have sufficiently dangerous capabilities.

A threshold-based approach like this has several virtues. It is very simple,
objective, determinable \emph{ex ante},\footnote{At least, determinable from the
planned specifications of the training run of an AI model, though of course
final FLOP usage will not be determined until the training run is complete.
However, AI developers tend to carefully plan the FLOP usage of training runs
for both technical and financial reasons.} and (due to the high price of
compute) is correlated with the ability of the developer to pay compliance
costs. One drawback, however, is that the same number of FLOP will produce
greater capabilities over time due to algorithmic improvements
\parencite{hernandez2020measuring}.  This means that, all else equal, the
probability that a foundation model below the threshold will have sufficiently
dangerous capabilities will increase over time.  These problems may not be
intractable. For example, a FLOP threshold could formulaically decay over time
based on new models’ performance on standardized benchmarks, to attempt to
account for anticipated improvements in algorithmic efficiency.\footnote{As an
analogy, many monetary provisions in US law are adjusted for inflation based on
a standardized measure like the consumer price index.}

A related approach could be to define the regulatory target by reference to the
most capable broad models that have been shown not to have sufficiently
dangerous capabilities. The idea here is that, if a model has been shown not to
have sufficiently dangerous capabilities, then every model that can be expected
to perform worse than it should also not be expected to have sufficiently
dangerous capabilities. Regulation would then apply only to those models that
exceed the capabilities of models known to lack sufficiently dangerous
capabilities. This approach has the benefit of updating quickly based on
observations from newer models. It would also narrow the space of regulated
models over time, as regulators learn more about which models have sufficiently
dangerous capabilities.

However, this definition has significant downsides too. First, there are many
variables that could correlate with possession of dangerous capabilities, which
means that it is unclear ex ante which changes in development processes could
dramatically change capabilities. For example, even if model~A dominates
model~B on many obvious aspects of its development (e.g., FLOP usage, dataset
size), B may dominate A on other important aspects, such as use of a new and
more efficient algorithm, or a better dataset. Accordingly, the mere fact that
a B is different from A may be enough to make B risky,\footnote{Compare the
definition of “frontier AI” used in \cite{shevlane2023model}: “models that are
both (a)~close to, or exceeding, the average capabilities of the most capable
existing models, and (b)~different from other models, either in terms of scale,
design (e.g. different architectures or alignment techniques), or their
resulting mix of capabilities and behaviours…”} unless the regulator can
carefully discriminate between trivial and risk-enhancing differences. The
information needed to make such a determination may also be highly sensitive
and difficult to interpret. Overall, then, determining whether a newer model
can be expected to perform better than a prior known-safe model is far from
straightforward.

Another potential problem with any compute-based threshold is that models below
it could potentially be open-sourced and then further trained by another actor,
taking its cumulative training compute above the threshold. One possible
solution to this issue could be introducing minimal requirements regarding the
open-sourcing of models trained using one or two orders of magnitude of compute
less than any threshold set. 

Given the uncertainty surrounding model capabilities, any definition will
likely be overinclusive. However, we emphasize the importance of creating broad
and clear ex ante exemptions for models that have no reasonable probability of
possessing dangerous capabilities. For example, an initial blanket exemption
for models trained with fewer than (say) 1E26 FLOP\footnote{Using public
FLOP per dollar estimates contained in~\cite{EpochAI:cost} (Epoch AI) and
\cite{lohn2022compute}, this would cost nearly or more than \$100 million in
compute alone.} could be appropriate, to remove any doubt as to whether such
models are covered. Clarity and definitiveness of such exemptions is crucial to
avoid overburdening small and academic developers, whose models will likely
contribute very little to overall risk.

\section{Scaling laws in Deep Learning}
\label{appendix:b}

This appendix describes results from the scaling laws literature which shape
the regulatory challenge posed by frontier AI as well as the available
regulatory options. This literature focuses on relationships between measures
of model performance (such as test loss) and properties of the model training
process (such as amounts of data, parameters, and compute). Results from this
literature of particular relevance to this paper include: (i)~increases in the
amount of compute used to train models has been an important contributor to AI
progress; (ii)~even if the increase in compute starts contributing less to
progress, we still expect frontier AI models to be trained using large amounts
of compute; (iii)~though scale predictably increases model performance on the
training objective, particular capabilities may improve or change unexpectedly,
contributing to the Unexpected Capabilities Problem.

In recent years, the Deep Learning Revolution has been characterized by the
considerable scaling up of the key inputs into neural networks, especially the
quantity of computations used to train a deep learning system (“compute”)
\parencite{Bashir2022}, as illustrated in Figure 4.

\begin{figure}
\begin{center}
\includegraphics[width=.9\textwidth]{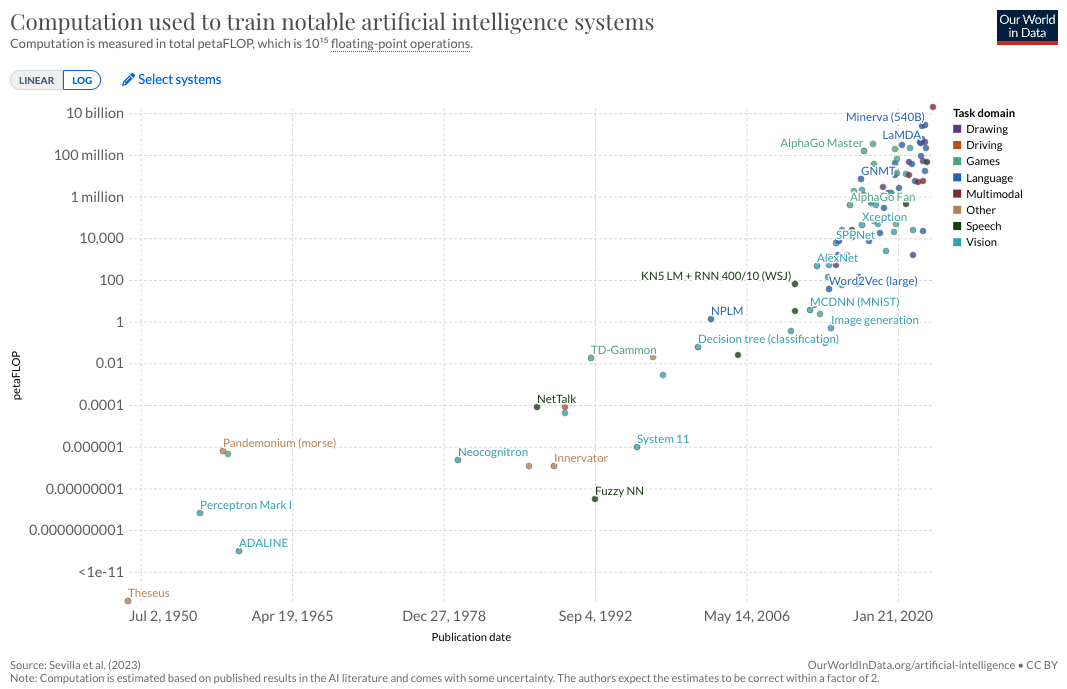}
\caption{Computation used to train notable AI systems.
Note logarithmic $y$-axis.
Source: \cite{OWID:computation} based on data from \cite{sevilla2022compute}.}
\end{center}
\label{figure:computation}
\end{figure}

Empirically, scaling training compute has reliably led to better performance on
many of the tasks AI models are trained to solve, and many similar downstream
tasks \parencite{Villalobos2023}.  This is often referred to as the “Scaling
Hypothesis”: the expectation that scale will continue to be a primary predictor
and determinant of model capabilities, and that scaling existing and
foreseeable AI techniques will continue to produce many capabilities beyond the
reach of current systems.\footnote{See \cite{Gwern:scaling, Sutton2019,
Bashir2022, bommasani2022opportunities}. For a skeptical take on the Scaling
Hypothesis, see \cite{lohn2022compute}.}

\begin{figure}
\begin{center}
\includegraphics[width=.9\textwidth]{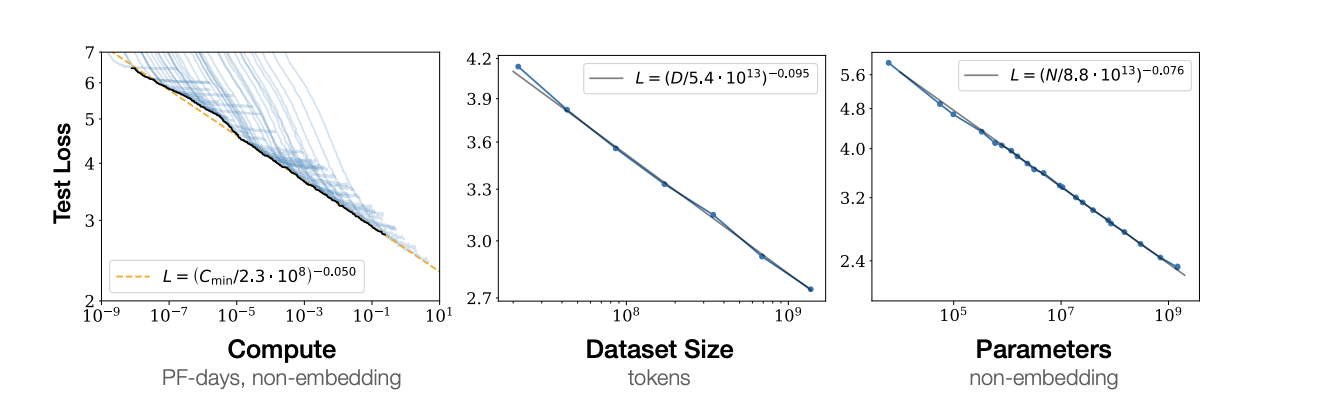}
\caption{Scaling reliably leading to lower test loss.
See \cite{kaplan2020scaling}.
The scaling laws from this paper have been updated by
\cite{hoffmann2022training}.}
\end{center}
\label{figure:loss}
\end{figure}

We expect the Scaling Hypothesis to account for a significant fraction of
progress in AI over the coming years, driving increased opportunities and
risks. However, the importance of scaling for developing more capable systems
may decrease with time, as per research which shows that the current rate of
scaling may be unsustainable \parencite{lohn2022compute, Heim2023,
Cottier2022}.

Even if increases in scale slow down, the most capable AI models are still
likely going to be those that can effectively leverage large amounts of
compute, a claim often termed “the bitter lesson” \parencite{Sutton2019}.
Specifically, we expect frontier AI models to use vast amounts of compute, and
that increased algorithmic efficiency \parencite{openai:efficiency} and data
quality \parencite{sorscher2023neural} will continue to be important drivers of
AI progress.

Scaling laws have other limits. Though scaling laws can, as illustrated in \chyperref[figure:loss]{Figure~5}, reliably predict the
loss of a model on its training objective – such as predicting the next word in
a piece of text – that is currently an unreliable predictor of downstream
performance on individual tasks. For example, tasks can see inverse scaling,
where scaling leads to worse performance \parencite{mckenzie2023inverse,
perez2022discovering, koralus2023humans}, though further scaling has overturned
some of these findings \parencite{openai2023gpt4}.

Model performance on individual tasks can also increase unexpectedly: there may
be “emergent capabilities” \parencite{Ganguli2022, schaeffer2023emergent}. Some
have argued that such emergent capabilities are a “mirage”
\parencite{schaeffer2023emergent}.  They argue that the emergence of
capabilities is primarily a consequence of how they are measured.  Using
discontinuous measures such as multiple choice answers or using an exact string
match, is more likely to “find” emergent capabilities than if using continuous
measures – for example, instead of measuring performance by exact string match,
you measure it based on proximity to the right answer.

We do not think this analysis comprehensively disproves the emergent
capabilities claim \parencite{Wei2023}.  Firstly, discontinuous measures are
often what matter. For autonomous vehicles, what matters is how often they
cause a crash. For an AI model solving mathematics questions, what matters is
whether it gets the answer exactly right or not. Further, even if continuous
“surrogate” measures could be used to predict performance on the discontinuous
measures, the appropriate choice of a continuous measure that will accurately
predict the true metric is often unknown a priori. Such forecasts instead
presently require a subjective choice between many possible alternatives, which
would lead to different predictions on the ultimate phenomenon. For instance,
is an answer to a mathematical question “less wrong” if it’s numerically closer
to the actual answer, or if a single operation, such as multiplying instead of
dividing, led to an incorrect result?

Nevertheless, investing in further research to more accurately predict
capabilities of AI models ex ante is a crucial enabler for effectively
targeting policy interventions, using scaling laws or otherwise.
\end{appendices}

\printbibliography
\end{document}